\documentclass[a4paper,UKenglish,cleveref,
autoref]{lipics-v2021}
\usepackage{tikz,float,accents,url}
\usepackage[linguistics]{forest}
\usetikzlibrary{graphs}
\usetikzlibrary{arrows.meta}

\pdfoutput=1
\hideLIPIcs

\title{
A pumping-like lemma
for languages over infinite alphabets}

\titlerunning{Pumping lemma}

\author{Yoav {Danieli}}{Technion -- Israel Institute of Technology, Faculty of Computer Science, Haifa, Israel.}
{yoavd@campus.technion.ac.il}{https://orcid.org/0000-0001-8774-064X}{}

\authorrunning{Y. Danieli}

\Copyright{Yoav Danieli}



\nolinenumbers
\makeatletter
\AtBeginDocument{\let\@oddfoot\@empty\let\@evenfoot\@empty}
\makeatother

\begin{document}

\newcommand{\trans}[1]{ \xrightarrow{#1}  }
\newcommand{\transword}[1]{ \xrightarrow{#1}\mathrel{\vphantom{\to}^*}}

\newcommand{\bA}{\mbox{\boldmath$A$}}
\newcommand{\bB}{\mbox{\boldmath$B$}}
\newcommand{\bC}{\mbox{\boldmath$C$}}
\newcommand{\bG}{\mbox{\boldmath$G$}}
\newcommand{\bL}{\mbox{\boldmath$L$}}
\newcommand{\bS}{\mbox{\boldmath$S$}}
\newcommand{\bX}{\mbox{\boldmath$X$}}
\newcommand{\bY}{\mbox{\boldmath$Y$}}

\newcommand{\N}{{\mathbb N}}

\newcommand{\bN}{\mbox{\boldmath$N$}}
\newcommand{\bP}{\mbox{\boldmath$P$}}
\newcommand{\bT}{\mbox{\boldmath$T$}}
\newcommand{\ba}{\mbox{\boldmath$a$}}
\newcommand{\bc}{\mbox{\boldmath$c$}}
\newcommand{\bd}{\mbox{\boldmath$d$}}
\newcommand{\bj}{\mbox{\boldmath$j$}}
\newcommand{\f}{\mbox{\boldmath$f$}}
\newcommand{\id}{\mbox{\boldmath$id$}}
\newcommand{\bl}{\mbox{\boldmath$l$}}
\newcommand{\nl}{\mbox{\boldmath$nl$}}
\newcommand{\br}{\mbox{\boldmath$r$}}
\newcommand{\bu}{\mbox{\boldmath$u$}}
\newcommand{\bv}{\mbox{\boldmath$v$}}
\newcommand{\bw}{\mbox{\boldmath$w$}}
\newcommand{\balpha}{\mbox{\boldmath$\alpha$}}
\newcommand{\bDelta}{\mbox{\boldmath$\Delta$}}
\newcommand{\bdelta}{\mbox{\boldmath$\delta$}}
\newcommand{\btheta}{\mbox{\boldmath$\theta$}}
\newcommand{\bsigma}{\mbox{\boldmath$\sigma$}}
\newcommand{\btau}{\mbox{\boldmath$\tau$}}
\newcommand{\bupsilon}{\mbox{\boldmath$\upsilon$}}
\newcommand{\bpsi}{\mbox{\boldmath$\psi$}}
\newcommand{\bphi}{\mbox{\boldmath$\varphi$}}
\newcommand{\bchi}{\mbox{\boldmath$\chi$}}
\newcommand{\bomega}{\mbox{\boldmath$\omega$}}
\newcommand{\bpi}{\mbox{\boldmath$\pi$}}
\newcommand{\BN}{{\mathbb{N}}}
\newcommand{\bQ}{{\mathbb{Q}}}

\newcommand{\mmm}[1]{\mbox{\boldmath \tiny$#1$}}
\newcommand{\mm}[1]{\mbox{\boldmath \scriptsize$#1$}}

\newcommand{\sba}{{\mm A}}
\newcommand{\sbC}{{\mm C}}
\newcommand{\sbc}{{\mm c}}
\newcommand{\sbG}{{\mm G}}
\newcommand{\sbphi}{{\mm \varphi}}
\newcommand{\sbtheta}{{\mm \theta}}
\newcommand{\sbpsi}{{\mm \psi}}
\newcommand{\sbtau}{{\mm \tau}}
\newcommand{\sbupsilon}{{\mm \upsilon}}
\newcommand{\sbdelta}{{\mm \delta}}
\newcommand{\sbsigma}{{\mm \sigma}}
\newcommand{\sbchi}{{\mm \chi}}
\newcommand{\sbomega}{{\mm \omega}}
\newcommand{\sbsharp}{{\mm \#}}

\newcommand{\tbc}{{\mmm c}}
\newcommand{\C}{{\cal C}}
\newcommand{\F}{{\cal F}}
\newcommand{\sbw}{\mm w}

\newcommand{\tba}{{\mmm A}}

\newcommand{\sbu}{\sbw_0}
\newcommand{\tbw}{\mmm w}
\newcommand{\tbu}{\tbw_0}

\newcommand{\tbsigma}{{\mmm \sigma}}
\newcommand{\tbphi}{{\mmm \varphi}}
\newcommand{\tbtau}{{\mmm \tau}}
\newcommand{\tbupsilon}{{\mmm \upsilon}}

\renewcommand{\lambda}{\mu}
\renewcommand{\star}{{\texttt{action}}}

\begin{titlepage}

\title{A pumping-like lemma for
languages
\\
over infinite alphabets}

\maketitle
\thispagestyle{empty}

\begin{abstract}
We prove a pumping-like lemma for languages accepted by one-register
alternating finite-memory automata over an infinite alphabet.
Every
sufficiently long accepted word has a nonempty factor in a bounded suffix
that can be repeated under successive powers of a single alphabet
permutation.
The proof analyzes accepting runs backwards, counting register values by
their state profiles within the remaining suffix and summarizing values
outside it.

We derive effective
semi-linearity of the sets of accepted word lengths and prove non-closure
under reversal, concatenation, and Kleene star. We also show that this pumping lemma
cannot be extended to alternating automata
with two registers.
\end{abstract}

\end{titlepage}

\addtocounter{page}{-1}

\section{Introduction}
\label{s: introduction}

{\em Finite-memory automata}~\cite{KaminskiF90,KaminskiF94}
generalize
classical Rabin-Scott finite-state automata~\cite{RabinS59} to
{\em infinite} alphabets.
By equipping the automata with a finite set
of registers (called in~\cite{KaminskiF90,KaminskiF94}
{\em windows}),
which store letters from the infinite alphabet
during computation, and
restricting the power of the automaton to
comparing the input letters with register contents
and copying input letters to registers,
the automaton retains a fixed finite-memory of input letters.
Consequently, the languages accepted by
finite-memory automata possess properties
similar to regular languages and are,
thus, termed {\em quasi-regular} languages.

An important facet of finite-memory automata is a certain
{\em indistinguishability}
view of the infinite alphabet embedded in
the modus operandi of the automaton.
The language accepted by an automaton is invariant under
permutations fixing the initial
assignment and the constants.
Thus, the actual symbols occurring in the input are of
no real significance.
Only the initial and repetition patterns matter.

Automata over infinite alphabets have gained increasing importance
in computer science as they provide formal models
for analyzing systems that operate over unbounded data domains.
Such models are essential for specifying and verifying properties
of XML documents, database queries,
and programs with variables over infinite domains.
The ability to reason about infinite alphabets while maintaining
decidability of key properties makes these models particularly
valuable for the formal verification of data-aware systems.

Over the years,
many models of automata over infinite alphabets
have been proposed
(see surveys in \cite{Segoufin06,Kara16,ChenSW16}),
though most of these models are incomparable.
In the absence of a definitive model
for managing infinite alphabets,
evaluating a formalism requires consideration
of its desirable properties, including:
expressive power, closure and regular properties,
decidability and complexity of classical problems, and
applicability of the model.

Finite-memory automata provide a structured way to reason
about such systems by focusing on patterns and repetitions
of data values rather than their specific identities.
This perspective is particularly useful in applications
where the exact values of data are less important
than their relative behavior over time.
Note that quasi-regular languages are a particular case of
nominal sets
(also known as sets with atoms or Fraenkel-Mostowski sets)~\cite{Bojanczyk19},
and finite-memory automata themselves are expressively
equivalent to nominal automata
(also known as orbit-finite automata)~\cite{BojanczykKL14}.

However,
the non-emptiness problem for finite-memory automata
is NP-complete~\cite{SakamotoI00} and
the universality problem is undecidable for
automata with two (or more) registers~\cite{NevenSV04}.

Nevertheless,
when restricted to a single register,
the emptiness problem is decidable
for the {\em alternating} variant of
finite-memory automata~\cite{DemriL09,FigueiraHL16,GenkinKP14},
albeit with non-primitive recursive complexity~\cite{DemriL09}.
Since alternation extends the power of nondeterministic
automata by introducing existential and universal states,
effectively allowing the automaton to explore multiple
computation paths in parallel~\cite{ChandraKS81},
the universality and containment problems
are decidable for these automata as well.

The restriction of finite-memory automata to a single register
(while still allowing a finite set of constant symbols)
is of significance:
its Parikh image is rational and has star-height
at most two~\cite{DanieliLICS26},
the existence of a deterministic equivalent automaton
is decidable for a fixed number of registers \cite{ClementeLP22}, and,
for ordered alphabets,
the intersection of nondeterministic
and co-nondeterministic one-register languages
lies within the deterministic class \cite{KlinlT21}.

This paper deals with alternating
finite-memory automata with one register.
Such automata are tightly related to
the linear and branching temporal
logics with the freeze quantifier.
In particular,
the satisfiability problem for formulas
in linear temporal logic
with forward operators and one
freeze quantifier is decidable,
see~\cite{DemriL09}.
In addition, these models are particularly robust
and possess many regular properties,
being closed under Boolean operations
and subsuming various other automata models,
such as top-view pebble automata
and weak two-pebble automata~\cite{Tan10}.
Recently,
it was shown in~\cite{DanieliK25} that the boundness problem
for these models is also decidable, by proving an
extension lemma for sufficiently long words.
Also,
there is a strong connection
to alternating timed automata with one clock,
such that complexity and decidability
results transfer back and forth
between these models \cite{FigueiraHL16}.

A fundamental property of classical finite-state automata
is their regularity,
which is often associated with the pumping lemma.
The pumping lemma serves as a crucial tool for proving
that certain languages are not regular by identifying
structural constraints on words within the language.

For general many-register finite-memory automata,
it is well known that the classical pumping lemma
does not necessarily hold,
see~\cite[Example~3]{KaminskiF94}.
Therefore, in any suitable pumping lemma for
languages over infinite alphabets, the pumped patterns are not, necessarily,
identical, but, instead, are equivalent modulo some permutation
of the infinite alphabet.
A pumping lemma of this kind has been proven for
finite-memory automata in~\cite{NakanishiTS23},
though it does not involve the action of permutations
on repeated subwords
and only establishes the existence of such pumped patterns.

The main result of this paper is a kind of pumping lemma
for languages accepted by
one-register alternating finite-memory automata.
Its applications include
effective semi-linearity of the sets of accepted word lengths and
non-closure of this language class under reversal, concatenation, and
Kleene star.
A language accepted with two registers has a non-semilinear
length set, showing that the pumping property does not extend to that
setting.

This paper is organized as follows.
In the next section, we recall the definition of
finite-memory automata from~\cite{KaminskiF94} and,
in~\Cref{s: pumping lemma for fma},
we prove a pumping lemma for these automata.
Sections~\ref{s: afma} and~\ref{s: pumping lemma}, respectively,
contain the definition of alternating finite-memory automata and
the statement of the pumping lemma for these automata,
that is the main result of our paper.
In \Cref{s: app}, we present some applications of the pumping lemma,
including effective semi-linearity and the three non-closure results.
\Cref{s: pf roadmap} contains the proof of our pumping lemma.
Finally,~\Cref{s: remark} contains some remarks concerning
the complexity of computing
the pumping lemma parameter and
possible extensions for ordered alphabets and
timed automata.

\section{Finite-memory automata}
\label{s: fma}

Throughout this paper, we employ the following conventions.
\begin{itemize}
\item
$\Sigma$ is a fixed infinite alphabet.
\item
Symbols in $\Sigma$ are denoted by $\sigma$,\,$\theta$, etc.,
sometimes indexed or primed.
\item
Bold lowercase Greek letters $\bsigma$,\,$\btheta$, etc.,
also sometimes indexed or primed,
denote words over $\Sigma$.
\item
Symbols that occur in a word denoted by a boldface letter
are always denoted by
the same \emph{non-boldface} letter with an appropriate subscript.
Thus $\sigma_h$ denotes the letter at position $h$ of $\bsigma$.
Above word-transition arrows, word symbols are printed in small,
non-bold type.
\item
For word
$\bsigma = \sigma_1 \sigma_2 \cdots \sigma_n \in \Sigma^\ast$,
we denote by $[\bsigma]$
the set of letters occurring in $\bsigma$:
\[
[\bsigma] = \{ \sigma_1 , \sigma_2 , \ldots , \sigma_n \}
\]
and
call this set the {\em contents} of $\bsigma$.
\item
For a subset $\Phi$ of $\Sigma$ and a positive integer $r$,
we denote by $\Phi^{r_{\neq}}$ the subset of $\Phi^r$
whose elements do not contain repeated letters:
\[
\Phi^{r_{\neq}} =
\left\{
\sigma_1\ldots\sigma_{r} \in \Phi^{r}
:
\mbox{for all} \ i \ \mbox{and} \ j, \
\mbox{such that} \ i \neq j , \ \sigma_i \neq \sigma_j
\right\}
\,
.
\]
\item The length of a word $\bsigma\in \Sigma^\ast$
is denoted by $|\bsigma|$ and the cardinality of a finite set $X$
is denoted by $|X|$.
\end{itemize}

Next, we recall the definition of finite-memory automata
from~\cite{KaminskiF90,KaminskiF94} in which
a fixed finite number of distinct letters
can be stored in the automaton memory during a computation.

\begin{definition}[{Cf.~\cite[Definition 1]{KaminskiF94}}]
\label{d: fma}
An \emph{$r$-register finite-memory automaton} (over $\Sigma$)
is a system
$\bA =
\langle S,s_0,F,\Delta,\btheta_0,
\lambda_\Delta , \lambda_= ,
\lambda_{\neq_{\text{skip}}},
\lambda_{\neq_{\text{replace}}} \rangle$
whose components are as follows.
\begin{itemize}
\item
$S$ is a finite set of \emph{states}.
\item
$s_0\in S$ is the \emph{initial state}.
\item
$F\subseteq S$ is the set of \emph{accepting states}.
\item
$\Delta \subset \Sigma$ is a finite set
of {\em distinguished} letters (constants).
\item
$\btheta_0 \in (\Sigma \setminus \Delta)^{r_{\neq}}$
is the {\em initial register assignment}.\footnote{
We require that the memory does not repeat letters,
as retaining a letter more than once is unnecessary.
Moreover, distinguished symbols from $\Delta$
cannot be stored in the registers,
as they are recognized without requiring allocated space.
}
\item
$\lambda_\Delta : S \times \Delta \rightarrow 2^S$,
$\lambda_= : S \times \{ 1,2,\ldots,r \} \rightarrow 2^S$,
$\lambda_{\neq_{\text{skip}}} : S \rightarrow 2^S$,
and
$\lambda_{\neq_{\text{replace}}}
:
S \rightarrow 2^{S \times \{ 1,2,\dots,r \}}$
are the \emph{transition} functions.

The intuitive meaning of these functions is as follows.
Let $\bA$ read a letter $\sigma$ being in
state $s$ with a letter $\theta_i$ stored in the $i$th register,
$i = 1,2,\ldots,r$.
\begin{itemize}
\item
If $\sigma \in \Delta$,
then $\bA$ enters a state from
$\lambda_\Delta(s,\sigma)$, with the same letters in the registers.
\item
If for some $i = 1,2,\ldots, r$, $\sigma = \theta_i$,
then $\bA$ enters a state from
$\lambda_=(s,i)$, with the same letters in the registers.
\item

If
$\sigma \not \in
\Delta \cup \{ \theta_1,\theta_2,\ldots,\theta_r \}$,
then,
either $\bA$ enters a state from
$\lambda_{\neq_{\text{skip}}}(s)$,
with the same letters in the registers, or
for some $(s^\prime,i) \in \lambda_{\neq_{\text{replace}}}(s)$,
$\bA$ enters $s^\prime$ and replaces, in the $i$th register,
$\theta_i$ with $\sigma$, leaving all other registers intact.
\end{itemize}
\end{itemize}
\end{definition}

A {\em configuration} of $\bA$ is a pair
in $S \times \left( \Sigma \setminus \Delta \right)^{r_{\neq}}$,
where the state component of the pair is
the current state and
the letters' components are the letters stored in the registers.
The configuration $(s_0,\btheta_0)$ is the
{\em initial} configuration, and
the configurations with the first component in $F$
are {\em accepting} (or {\em final}) configurations.

For configurations
$(s,\btheta)$ and $(s^\prime,\btheta^\prime)$,
where
$\btheta = \theta_1 \theta_2 \cdots \theta_r$
and
$\btheta^\prime =
\theta_1^\prime \theta_2^\prime \cdots \theta_r^\prime$,
we write
$(s,\btheta)
\trans{\sigma}
(s^\prime,\btheta^\prime)$,
if one of the following holds.
\begin{itemize}
\item
$\sigma \in \Delta$,
$s^\prime \in \lambda_\Delta(s,\sigma)$, and
$\btheta^\prime = \btheta$.
\item
For some $i = 1,2,\ldots,r$,
$\sigma = \theta_i$,
$s^\prime \in \lambda_=(s,i)$, and $\btheta^\prime = \btheta$.
\item
$\sigma \notin \Delta \cup [\btheta]$ and
either $s^\prime \in \lambda_{\neq_{\text{skip}}}(s)$,
and $\btheta^\prime = \btheta$,
or for some $i = 1,2,\ldots,r$,
$(s^\prime,i) \in \lambda_{\neq_{\text{replace}}}(s)$,
and, for $j = 1,2,\ldots,r$,
\[
\theta_j^\prime =
\begin{cases}
\theta_j & \mbox{if} \, j \neq i.
\\
\sigma & \mbox{if} \,  j = i.
\end{cases}
\]
\end{itemize}

A run of $\bA$ on a word
$\bsigma = \sigma_1\sigma_2\cdots\sigma_{n}$
is a sequence of configurations
$c_0,c_1,\ldots,c_n$ such that for all $i = 1,2,\ldots,n$,
$c_{i-1} \trans{\sigma_i} c_i$.
In such a case, we shall also write
$c_0 \transword{\sigma} c_n$.

A run is {\em accepting},
if it starts from the {\em initial}
configuration $(s_0,\btheta_0)$ and
ends in a final configuration.
A word is accepted by $\bA$,
if there exists an accepting run of $\bA$ on it.
The language of all accepted words
is denoted by $L \left( \bA \right) $.

\begin{example}[{\cite[Example 1]{KaminskiF94}}]
\label{e: two}
Consider the one-register finite-memory automaton $\bA$
shown in~\Cref{fig: aut l diff}.\footnote{
Since $r = 1$, the register component of the transition functions
is omitted.}

\begin{figure}[htbp]
\centering
\begin{tikzpicture}[scale=0.15]
\tikzstyle{every node}+=[inner sep=0pt]
\draw [black] (6.4,-28.21) -- (15.35,-28.29);
\fill [black] (15.35,-28.29) -- (14.55,-27.78) -- (14.55,-28.78);
\draw [black] (18.4,-28.2) circle (3);
\draw (18.4,-28.2) node {$s_0$};
\draw [black] (39.2,-28.3) circle (3);
\draw (39.2,-28.3) node {$s$};
\draw [black] (60.3,-28.2) circle (3);
\draw (60.3,-28.2) node {$f$};
\draw [black] (60.3,-28.2) circle (2.4);
\draw [black] (17.077,-25.52) arc (234:-54:2.25);
\draw (18.4,-20.95) node [above] {$=,\neq_{\text{skip}}$};
\fill [black] (19.72,-25.52) -- (20.6,-25.17) -- (19.79,-24.58);
\draw [black] (21.4,-28.21) -- (36.2,-28.29);
\fill [black] (36.2,-28.29) -- (35.4,-27.78) -- (35.4,-28.78);
\draw (28.8,-28.75) node [below] {$=,\neq_{\text{replace}}$};
\draw [black] (42.2,-28.29) -- (57.3,-28.21);
\fill [black] (57.3,-28.21) -- (56.5,-27.72) -- (56.5,-28.72);
\draw (49.75,-29.46) node [below] {$=$};
\draw [black] (37.877,-25.62) arc (234:-54:2.25);
\draw (39.2,-21.05) node [above] {$\neq_{\text{skip}}$};
\fill [black] (40.52,-25.62) -- (41.4,-25.27) -- (40.59,-24.68);
\draw [black] (58.977,-25.52) arc (234:-54:2.25);
\draw (61.3,-20.95) node [above] {$=,\neq_{\text{skip}}$};
\fill [black] (61.62,-25.52) -- (62.5,-25.17) -- (61.69,-24.58);
\end{tikzpicture}
\caption{The graph representation of $\bA$.}
\label{fig: aut l diff}
\end{figure}
It is easy to see that the language of $\bA$ consists of
all words over $\Sigma$
in which some letter appears more than once:
\[
L(\bA) =
\{ \sigma_1 \sigma_2 \cdots \sigma_n : \ \mbox{there exist} \
1 \leq i < j \leq n \ \mbox{such that} \ \sigma_i = \sigma_j \} .
\]
The initial register value is arbitrary for this language. The loop at
$s_0$ skips the current letter whether or not it equals that value.
When
the automaton chooses the first occurrence of a repeated letter, it moves
to $s$ and retains the register on equality or replaces it otherwise.
Thus it can remember any selected input letter, independently of the
initial assignment.
\end{example}

\begin{example}
\label{e: l first}
The language
\[
L_{\text{first}}=
\{ \sigma_1 \sigma_2 \cdots \sigma_n\in\Sigma^+ : \ \mbox{for all} \
1 < i  \leq n, \ \sigma_1 \neq \sigma_i \}
\]
is accepted by the automaton $\bB$
in \Cref{fig: aut l first}.

\begin{figure}[htbp]
\centering
\begin{tikzpicture}[scale=0.15]
\tikzstyle{every node}+=[inner sep=0pt]
\draw [black] (6.4,-28.21) -- (15.35,-28.29);
\fill [black] (15.35,-28.29) -- (14.55,-27.78) -- (14.55,-28.78);
\draw [black] (18.4,-28.2) circle (3);
\draw (18.4,-28.2) node {$s_0$};
\draw [black] (39.2,-28.3) circle (3);
\draw (39.2,-28.3) node {$f$};
\draw [black] (39.2,-28.2) circle (2.4);
\draw [black] (21.4,-28.21) -- (36.2,-28.29);
\fill [black] (36.2,-28.29) -- (35.4,-27.78) -- (35.4,-28.78);
\draw (28.8,-28.75) node [below] {$=,\neq_{\text{replace}}$};
\draw [black] (37.877,-25.62) arc (234:-54:2.25);
\draw (39.2,-21.05) node [above] {$\neq_{\text{skip}}$};
\fill [black] (40.52,-25.62) -- (41.4,-25.27) -- (40.59,-24.68);
\end{tikzpicture}
\caption{The graph representation of $\bB$.}
\label{fig: aut l first}
\end{figure}

\end{example}

The pumping lemmas in this paper involve the following definitions.

\begin{definition}
\label{d: permutation}
The order of a permutation $\alpha : \Sigma \rightarrow \Sigma$
is the smallest positive integer $k$ such that
$\alpha^k$ is the identity permutation:
$\alpha^{k} = \id$,\footnote{
As usual, the product of two permutations is their composition
and powers are defined recursively by
$\alpha^k = \alpha \circ \alpha^{k-1}$.}
if no such $k$ exists, the order of $\alpha$
is infinite.
Moreover, we say that $\alpha$ is a $\Delta$-{\em permutation}
if it is invariant on $\Delta$
(i.e., for all $\delta\in\Delta,\alpha(\delta)=\delta$).
\end{definition}

\begin{definition}
\label{d: extension}
Let $\alpha$ be a permutation of $\Sigma$.
We extend $\alpha $ to configurations by
$\alpha(s,\btheta) = (s,\alpha(\btheta))$
and then to finite sets of configurations by
$\alpha(C) = \{\alpha(c) : c \in C \}$.
\end{definition}

\begin{proposition}[Equivariance of FMA]
\label{p: permutation word fma}

Let $\alpha$ be a $\Delta$-permutation of $\Sigma$,
$\bsigma \in \Sigma^\ast$, and
let $c$ and $c^\prime$ be configurations of $\bA$ such that
$c \transword{\sigma} c^\prime$.
Then
$\alpha(c)
\transword{\alpha(\sigma)}
\alpha(c^\prime)$.
\end{proposition}

The proof of~\Cref{p: permutation word fma}
is similar to that of~\cite[Lemma~1]{KaminskiF94} and is omitted.

\begin{remark}
The same model also handles \emph{data words} over $H\times\Sigma$,
where $H$ is a finite set of labels.
Take a disjoint copy of $H$ as constants
and encode $\langle h_1,a_1\rangle\langle h_2,a_2\rangle\cdots
\langle h_n,a_n\rangle$ by
$h_1a_1h_2a_2\cdots h_na_n$.
Finite control tracks the alternating
label and data positions, while registers store and compare the
data letters.
\end{remark}

\section{Pumping lemma for finite-memory automata}
\label{s: pumping lemma for fma}

It has been shown in~\cite[Example~3]{KaminskiF94} that
the pumping lemma for regular languages~\cite[Lemma~8]{RabinS59}
does not hold for the quasi-regular ones.
A variant of the pumping lemma
for quasi-regular languages
was later published in~\cite{NakanishiTS23},
although it had previously been obtained
independently by two groups \cite{Klin14}.\footnote{The author
in \cite{Klin14}
also posed the question of
whether their pumping lemma
extends to alternating automata.
However, it does not apply directly
to this setting;
see the discussion in \Cref{s: pl examples}.}
The pumping lemma presented below
refines these results in which
\begin{itemize}
\item
the pumped patterns are generated by a
fixed finite-order permutation, and
\item
an explicit upper bound is given on the order of that permutation.
\end{itemize}

\begin{theorem}
\label{t: Pumping_Lemma_2}
Let
$\bA =
\langle S,s_0,F,\Delta,\btheta_0,
\lambda_\Delta , \lambda_= ,
\lambda_{\neq_{\text{skip}}} ,
\lambda_{\neq_{\text{replace}}} \rangle$
be an $r$-register finite-memory automaton.
Then for every word $\bsigma \in L(\bA)$
and every subword $\bchi$ of $\bsigma$ with
$\bsigma = \bpsi \bchi \bomega$ and
$|\bchi|>|S|$,
there exists a decomposition $\bchi = \btau \bupsilon \bphi$
and
a $\Delta$-permutation $\alpha$ of order at most $(2r)!$
such that
\begin{itemize}
\item
$0<|\bupsilon|\leq|S|$,
\item
for all $k = 1,2,\ldots,$
\begin{equation}
\label{eq: k fma}
\bpsi\btau\bupsilon
\alpha \left(\bupsilon\right)
\alpha^2 \left(\bupsilon\right)
\cdots
\alpha^k \left(\bupsilon\right)
\alpha^k \left(\bphi\bomega\right)
\in L(\bA)
\,
,
\end{equation}
\item and
\begin{equation}
\label{eq: zero}
\bpsi \btau
\alpha^{-1} \left( \bphi \bomega \right) \in L(\bA)
\,
.
\end{equation}
\end{itemize}
\end{theorem}

\begin{proof}
Fix an accepting run $(s_h,\btheta_h)_{0\leq h\leq m}$
on $\bsigma$. The pigeonhole principle gives
\begin{equation*}
\label{eq: repeated-state}
|\bpsi|\leq i<j\leq|\bpsi|+|S|,
\qquad s_i=s_j.
\end{equation*}
Choose $\bchi=\btau\bupsilon\bphi$ with
$|\btau|=i-|\bpsi|$ and $|\bupsilon|=j-i$.
Then $0<|\bupsilon|\leq|S|$ and
\begin{equation}
\label{eq: selected-fma-run}
(s_0,\btheta_0)
\transword{\psi\tau}(s_i,\btheta_i)
\transword{\upsilon}(s_i,\btheta_j)
\transword{\varphi\omega}(s_m,\btheta_m),
\qquad s_m\in F.
\end{equation}

Extend the coordinate-wise bijection between the register
assignments to a permutation of $[\btheta_i]\cup[\btheta_j]$
by closing its paths into cycles, and fix all other letters.
Since register letters avoid $\Delta$, this defines a
$\Delta$-permutation $\alpha$ satisfying
\begin{equation}
\label{eq: register-permutation}
\alpha(\btheta_i)=\btheta_j.
\end{equation}
By~\Cref{p: permutation word fma},
\eqref{eq: selected-fma-run} and~\eqref{eq: register-permutation}
yield
\begin{equation}
\label{eq: middle}
(s_i,\alpha^h(\btheta_i))
\transword{\alpha^h(\upsilon)}
(s_i,\alpha^{h+1}(\btheta_i))
\qquad(h\geq0),
\end{equation}
and
\begin{equation}
\label{eq: end}
(s_i,\alpha^{k+1}(\btheta_i))
\transword{\alpha^k(\varphi\omega)}
(s_m,\alpha^k(\btheta_m))
\qquad(k\geq-1).
\end{equation}
For $k\geq1$, concatenate the prefix run
in~\eqref{eq: selected-fma-run}, the runs
in~\eqref{eq: middle} for $0\leq h\leq k$,
and~\eqref{eq: end}, proving~\eqref{eq: k fma}.
Appending~\eqref{eq: end} with $k=-1$ directly to that
prefix proves~\eqref{eq: zero}.

Finally, $\alpha$ is supported on at most $2r$ letters,
so its order is at most $(2r)!$.\footnote{
In fact, the order of $\alpha$ is bounded by $g(2r)$
where $g$ is Landau's function~\cite{Landau03}.
Actually, it is possible to
construct an explicit permutation of order at most
$\max\{r+1,g(r)\}$, see~\cite[Theorem 14]{DanieliAfl26}.}
\end{proof}

Even though the classical pumping lemma does not hold for
quasi-regular languages, as~\Cref{c: periodicity} shows below,
some sufficiently long words in
a quasi-regular language possess periodicity properties.

\begin{corollary}
\label{c: periodicity}
If $L$ is an unbounded quasi-regular language,
then there are words
$\bsigma^\prime,
\bsigma^{\prime\prime}$, and
$\bsigma^{\prime\prime\prime}$
such that $|\bsigma^{\prime\prime}|>0$ and
for all $k = 0,1,\ldots$,
\[
\bsigma^\prime
\left( {\bsigma^{\prime\prime}}\right)^k
\bsigma^{\prime\prime\prime}
\in
L
\,
.
\]
\end{corollary}

\begin{proof}
Let $\bsigma\in L$ be a sufficiently
long word.
Using the notation of~\Cref{t: Pumping_Lemma_2},
$\bsigma^\prime = \bpsi\btau\bupsilon$,
$\bsigma^{\prime\prime} =
\alpha(\bupsilon)
\alpha^2(\bupsilon)
\cdots \alpha^{i}(\bupsilon)$,
and
$\bsigma^{\prime\prime\prime} =
\alpha^i(\bphi\bomega)=\bphi\bomega$,
where $i$ is the order of $\alpha$.
\end{proof}

\section{Alternating finite-memory automata}
\label{s: afma}

In this section, we recall the definition of
alternating finite-memory automata and
state some of its basic properties.
We restrict ourselves to one-register automata
with which we deal in this paper.

\begin{definition}
\label{d: afma}

An \emph{alternating one-register
finite-memory automaton} (over $\Sigma$)
is a system
$\bA = \langle S,s_0,F,\Delta,\theta_0,
\mu_\Delta,\mu_= ,\mu_{\neq} \rangle$
whose components are like in \Cref{d: fma} except that
the transition functions are as follows:
\begin{itemize}
\item
$\mu_\Delta : S \times \Delta \rightarrow 2^{2^S}$,
\item
$\mu_= : S \rightarrow 2^{2^S}$, and
\item
$\mu_{\neq} : S \rightarrow 2^{\left( 2^S \right)^2}$.
\end{itemize}

The intuitive meaning of these functions is as follows.
Let $\bA$ read a letter $\sigma$ being in
state $s$ with a letter $\theta$ stored in the register.
\begin{itemize}
\item
If $\sigma \in \Delta$, then,
for some $Q \in \mu_\Delta(s,\sigma)$,
the computation splits to the computations
from all states in $Q$
with $\theta$ in the register.
\item
If $\sigma = \theta$,

then, for some $Q \in \mu_= (s)$,
the computation splits to the computations
from all states in $Q$
with the same $\theta$ in the register.
\item

If $\sigma \notin \Delta \cup \{ \theta \}$,
then, for some $(Q^\prime,Q^{\prime\prime}) \in \mu_{\neq}(s)$,
the computation splits to
the computations from all states in $Q^\prime$
with $\theta$ in the register and
the computations from all states in $Q^{\prime\prime}$
with the current input letter $\sigma$ in the register.\footnote{
That is, the components $Q^\prime$ and $Q^{\prime\prime}$
of pair $(Q^\prime,Q^{\prime\prime})$ correspond to the values of
the transition functions
$\mu_{\neq_{\text{skip}}}$ and $\mu_{\neq_{\text{replace}}}$
from~\Cref{d: fma}, respectively.}
\end{itemize}
\end{definition}

\begin{remark}
The transition families in~\Cref{d: afma} encode positive Boolean
formulas in disjunctive normal form: choose a clause existentially,
then branch universally within it. For example,
$\{\{p\},\{q,r\}\}$ represents $p\lor(q\land r)$; $\emptyset$
and $\{\emptyset\}$ represent false and true.
For unequal input, $Q^\prime$ and $Q^{\prime\prime}$ separate
register-preserving and register-updating successors.
This convention is equivalent to explicit existential and universal
branching, preserving both the register count and the language.
\end{remark}

Like in the case of (non-alternating) finite-memory automata,
a {\em configuration} of $\bA$ is a pair in
$S \times (\Sigma \setminus \Delta)$.

The transition functions
$\mu_\Delta$, $\mu_= $,
and $\mu_{\neq}$ induce
the following transition function
$\mu^\sbc$ on a configuration $(s,\theta)$ and
an input letter $\sigma$, as follows,
\begin{itemize}
\item
If $\sigma \in \Delta$, then
$\mu^\sbc((s,\theta),\sigma) =
\{ Q \times \{ \theta \} : Q \in \mu_\Delta(s,\sigma) \}$.
\item
If $\sigma = \theta$, then
$\mu^\sbc((s,\theta),\sigma) =
\{ Q \times \{ \theta \} : Q \in \mu_= (s) \}  $.
\item
If $\sigma \not \in \Delta \cup \{ \theta \}$, then
$\mu^\sbc((s,\theta),\sigma) =
\{ Q^\prime \times \{ \theta \}
\cup Q^{\prime\prime} \times \{ \sigma \} :
(Q^\prime,Q^{\prime\prime}) \in \mu_{\neq}(s) \} $.
\end{itemize}

\begin{remark}
\label{r: finite word}
Note that $\mu^\sbc((s,\theta),\sigma)$ is a finite set of
finite sets of configurations.
\end{remark}

Next, we extend $\mu^\sbc$ to finite sets of configurations in
the standard ``alternating'' manner.
Let $C = \{ c_1,c_2,\ldots,c_k \}$ be a set of configurations,
$\sigma \in \Sigma$, and
let $\mu^\sbc(c_i,\sigma) =
\{ C_{i,1},C_{i,2},\ldots,C_{i,m_i} \}$,
$i = 1,2,\ldots,k$.
Then $\mu^\sbc(C,\sigma)$ consists of
all finite sets of configurations
of the form $\bigcup\limits_{i = 1}^k C_{i,j_i}$,
$j_i = 1,2,\ldots,m_i$.
That is, $C^\prime \in \mu^\sbc(C,\sigma)$, if there is a set
of finite sets of configurations
$\{ C_{1,j_1},C_{2,j_2},\ldots,C_{k,j_k} \}$,
$C_{i,j_i} \in \mu^\sbc(c_i,\sigma)$, $i = 1,2,\ldots,k$,
such that $C^\prime = \bigcup\limits_{i = 1}^k C_{i,j_i}$.
Like in the case of (non-alternating) finite-memory automata,
we write $C\trans{\sigma}C^\prime$
for $C^\prime \in \mu^{\sbc}\left(C,\sigma\right)$.

Now, we can define the notion of a {\em run} of $\bA$.
Let
$\bsigma = \sigma_1 \sigma_2 \cdots \sigma_n \in \Sigma^\ast$.
A {\em run}
of $\bA$ on $\bsigma$
is a sequence of finite sets of
configurations $C_0,C_1,\ldots,C_n$,
where
\begin{itemize}
\item
$C_0 = \{(s_0,\theta_0)\}$ and
\item
$C_i \trans{\sigma_{i+1}} C_{i+1}$,
$i = 0,1,\ldots,n - 1$.
\end{itemize}
That is
$C_{0}
\trans{\sigma_1}
C_1
\trans{\sigma_2}
C_2
\trans{\sigma_3}
\cdots
\trans{\sigma_{n}}C_{n} $
and,
like in the case of (non-alternating) finite-memory automata,
we also write $C_{0}\transword{\sigma}C_{n} $.
This run is {\em accepting}, if $C_n \subseteq F \times \Sigma$,
i.e., $C_n$ is a set of {\em accepting} configurations, and
the automaton {\em accepts} a word $\bsigma \in \Sigma^\ast$,
if there is an accepting run of $\bA$ on $\bsigma$.
The language $L(\bA)$
consists of all words accepted by $\bA$.

\begin{example}
\label{e: diff}
The language
$L_{\text{diff}}
=
\{ \sigma_1 \sigma_2 \cdots \sigma_n \in \Sigma^\ast :
\mbox{for all} \ i,j  \ \mbox{if} \ i\neq j, \mbox{then} \
\sigma_i \neq \sigma_j
\}
\, $,
that is the complement of the quasi-regular language $L$
from~\Cref{e: two},
is not quasi-regular,
because it is an unbounded language not containing words with periodic patterns,
in contradiction with  \Cref{c: periodicity}
(see also~\cite[Example 5]{KaminskiF94}).
An explicit one-register alternating automaton for $L_{\text{diff}}$
has states $S=\{q,p,d\}$, initial state $q$, accepting states
$F=\{q,p\}$, no distinguished letters, and any initial register value.
Its transitions are
\[
\begin{array}{c|c|c}
 s & \mu_=(s) & \mu_{\neq}(s)\\ \hline
 q & \{\{q,p\}\} & \{(\{q\},\{p\})\}\\
 p & \{\{d\}\} & \{(\{p\},\emptyset)\}\\
 d & \{\{d\}\} & \{(\{d\},\emptyset)\}
\end{array}
\]
The scanner $q$ continues and launches a checking branch $p$ for each
letter read. On an unequal input the new branch stores that input;
on equality it already holds the input value. Every checking branch
rejects if its stored letter reappears, by entering the nonaccepting
sink $d$. All branches must accept, so precisely the words with no
repeated letters are accepted, including the empty word.
\end{example}

\begin{example}
\label{e: subset}
Let
$L_{\subseteq}
=
\{
\bpsi \# \bomega :
\bpsi,\bomega \in L_{\text{diff}}
\ \mbox{and} \
[\bpsi] \subseteq [\bomega] \subset \Sigma \setminus \{ \# \}
\}$.
This language is accepted by an alternating
one-register finite-memory automaton.
The automaton verifies that the letter $\#$
appears exactly once in the input,
that the prefix before $\#$ and the suffix after $\#$
do not contain repeated letters,
and that every letter in the prefix also
appears in the suffix.
\end{example}

\begin{example}
\label{e: l last}
Let
$L_{\text{last}}=
\{ \sigma_1 \sigma_2 \cdots \sigma_n\in\Sigma^+ : \ \mbox{for all} \
1 \leq i  < n, \ \sigma_i \neq \sigma_n \}
\,
,$
that is the reversal of $L_{\text{first}}$ from \Cref{e: l first}.
A one-register alternating automaton recognizes it by modifying
the construction for $L_{\text{diff}}$ in~\Cref{e: diff}:
each checking branch only checks that its stored letter does not
occur at the final position of the remaining suffix.
It accepts if that suffix is empty.
Requiring the input to be nonempty gives precisely $L_{\text{last}}$.
\end{example}

\begin{proposition}
\label{p: permutation word}
Let $\alpha$ be a $\Delta$-permutation of $\Sigma$.
Then
$C \in \mu^\sbc(c,\sigma)$ implies
$\alpha(C) \in \mu^\sbc(\alpha(c),\alpha(\sigma))$.
\end{proposition}

The proof of~\Cref{p: permutation word}
is similar to that of
~\cite[Lemma~1]{KaminskiF94} and is omitted.
As a corollary, every run of the automaton
is mapped by permutations onto an equivalent run.

\begin{corollary}
\label{c: permutation run}
Let $C,C'$ be finite sets of configurations,
$\bsigma \in \Sigma^\ast$ be
such that
$C\transword{\sigma}C^\prime$ and
let $\alpha$ be a $\Delta$-permutation of $\Sigma$.
Then
$\alpha(C)
\transword{\alpha(\sigma)}
\alpha (C^\prime)$.
\end{corollary}

\section{Pumping lemma for alternating finite-memory automata}
\label{s: pumping lemma}

Our pumping lemma for alternating finite-memory automata
is as follows.

\begin{theorem}
\label{t: pumping_lemma}
Let $\bA$ be a one-register alternating finite-memory automaton.
There is a computable constant $N_\sba$ such that, for
every word $\bsigma \in L(\bA)$ that is longer than $N_\sba$,
there exists
a decomposition
$\bsigma = \btau \bupsilon \bphi$
of\, $\bsigma$ and
a $\Delta$-permutation
$\alpha$ of\, $\Sigma$ such that
\begin{itemize}
\item
$\left|\bupsilon\bphi\right|\leq N_\sba$,
\item
$\left|\bupsilon\right|>0$, and
\item
for all $k = 1,2,\ldots,$
\begin{equation}
\label{eq: pumping}
\btau\bupsilon
\alpha\left(\bupsilon\right)
\alpha^2\left(\bupsilon\right)\cdots
\alpha^{k}\left(\bupsilon\right)
\alpha^{k}\left(\bphi\right) \in L(\bA)
\,
.
\end{equation}
\end{itemize}
\end{theorem}

\subsection{
\texorpdfstring{\protect{\Cref{t: pumping_lemma}}}{The alternating pumping lemma}
vs.
\texorpdfstring{\protect{\Cref{t: Pumping_Lemma_2}}}{Theorem 7}
}
\label{s: pl examples}

Note that the statement of~\Cref{t: pumping_lemma} is weaker
than that of~\Cref{t: Pumping_Lemma_2}.
This is for the following three reasons, see the examples below
which illustrate the limitations of the classical approach
for languages over infinite alphabets.
\begin{itemize}
\item
The alphabet permutation is not necessarily of a finite order;
\item
pumping is not possible in every
sufficiently long pattern of $\bsigma$; and
\item
``shrinking'' the word as in~\eqref{eq: zero}
is not always possible.
\end{itemize}

First, we demonstrate the impossibility
of pumping certain patterns as they are,
without any modification.

\begin{example}
\label{e: diff pump}
The language $L_{\text{diff}}$ from~\Cref{e: diff} is unbounded,
but no word in it can be pumped in the classical sense,
as this would imply a letter repetition,
violating the language's defining property.
For the same reason, it follows from
\Cref{c: periodicity}
that no word in $L_{\text{diff}}$ can be pumped by any
finite order permutation of $\Sigma$.
\end{example}

The following (and a bit longer) example illustrates the necessity of
both the permutation and the restriction
to a bounded suffix.
General pumping with permutations in an arbitrary pattern,
like in~\Cref{t: Pumping_Lemma_2}, is not always possible.

\begin{example}
\label{e: subset pump}
The language $L_{\subseteq}$, from~\Cref{e: subset},
does not satisfy
the pumping lemma with permutations
applied arbitrarily within the word,
which can be shown as follows.

For $\bpsi = \psi_1\psi_2 \cdots \psi_m \in L_{\text{diff}}$,
the word
$\bsigma = \bpsi \# \bpsi^R$,
where $\bpsi^R = \psi_m\psi_{m-1} \cdots \psi_1$ is the reversal
of $\bpsi$,
belongs to $L_{\subseteq}$.

We claim that if a subword can be repeatedly
pumped by means of some permutation,
it must be contained within the suffix $\#\bpsi^R$ of $\bsigma$.

Assume to the contrary that there is a decomposition
$\bsigma = \bpsi \# \bpsi^{R} = \btau\bupsilon\bphi $
with
$|\bupsilon| > 0$ and a
$\Delta$-permutation $\alpha$ such that
$\bsigma_1 =
\btau\bupsilon\alpha (\bupsilon ) \alpha(\bphi) \in
L_{\subseteq} $
and
$\bupsilon$ is not a subword of the suffix
$\#\bpsi^{R}$ of $\bsigma$.

We distinguish between the cases of
$\alpha(\#) = \#$ and $\alpha(\#) \neq \#$.

In the former case,
$\bupsilon$ is a subword of the prefix $\bpsi$ of $\bsigma$,
because, otherwise,
in contradiction with the definition of $L_{\subseteq}$,
$\#$ would appear twice in
$\btau\bupsilon\alpha (\bupsilon ) \alpha(\bphi)$.
Thus, the prefix of
$\btau\bupsilon\alpha (\bupsilon ) \alpha(\bphi)$
up to $\#$ is longer than $\bpsi$,
which, again, contradicts the definition of $L_{\subseteq}$.

In the latter case,
the suffix $\alpha(\bupsilon)\alpha(\bphi)$ of $\bsigma_1$
contains the pattern $\alpha(\psi_{m}) \alpha(\#) \alpha(\psi_{m})$
in which two occurrences of $\alpha(\psi_m)$
are not separated by $\#$.
This, however, contradicts the definition of $L_{\subseteq}$.
\end{example}

The last example also shows that the pumping lemma
for alternating finite-memory automata cannot be strengthened
by the shrinking property~\eqref{eq: zero}.

\begin{example}
\label{e: shortenning}

For $\bsigma$ from~\Cref{e: subset pump},
the pumping can be applied only in the suffix
$\# \bpsi^R$ of $\bsigma$
and
the suffix of $\btau \alpha^{-1}(\bphi)$ beginning at $\#$
is shorter than $\bpsi$.
\end{example}

\section{Some applications of
\texorpdfstring{\protect{\Cref{t: pumping_lemma}}}{the alternating pumping lemma}
}
\label{s: app}

As the section title suggests,
this section contains a number of applications of~\Cref{t: pumping_lemma}.

\subsection{Disproving membership}
In this section, we apply~\Cref{t: pumping_lemma} to show that
a language is not accepted by
a one-register alternating finite-memory automaton.
Namely, we use the pumping lemma
to show three negative closure results on
the class of languages accepted by these automata.
These negative results are not unexpected, because runs of
alternating automata from universal states
are independent.\footnote{
For the same reason, the proof
of~\Cref{t: pumping_lemma} is rather involved.}

\begin{example}[Cf.~\Cref{e: subset}]
\label{e: supset}
The language
\[
L_\supseteq
=
\{ \bpsi \# \bomega :
\bpsi,\bomega \in L_{\text{diff}} , \
[\bpsi] \supseteq [\bomega] , \ \mbox{and} \
[\bpsi] , [\bomega] \subset \Sigma \setminus \{ \# \}
\}
\,
,
\]
where $L_{\text{diff}}$ is the language from~\Cref{e: diff},
is not accepted by a one-register alternating finite-memory automaton.
Consequently, the class of languages accepted by these automata is not closed under reversal.

Assume to the contrary that,
for some one-register alternating finite-memory automaton $\bA$,
$L_\supseteq = L(\bA)$.
Let $\bpsi \in L_{\text{diff}}$ be such that
$[\bpsi]\subset\Sigma\setminus\{\#\}$ and
$|\bpsi| > N_{\sba}$.
By the definition of $L_\supseteq$, it contains
$\bsigma=\bpsi\#\bpsi$.
Let $\bpsi\#\bpsi = \btau\bupsilon\bphi$,
$\btau = \btau^\prime\#\btau^{\prime\prime}$
be the decomposition of $\bsigma$ and
$\alpha$ be
the $\Delta$-permutation provided
by~\Cref{t: pumping_lemma}.
Then, by that theorem,
\[
\bsigma^\prime=\btau^\prime
\#\btau^{\prime\prime}
\bupsilon\alpha(\bupsilon)
\alpha(\bphi)
\in
L_\supseteq
\,
,
\]
notice the following length comparison between
the prefix and suffix of the word $\bsigma^\prime$,
\begin{equation}
\label{eq: len comp}
|\btau^{\prime\prime}\bupsilon\alpha(\bupsilon)\alpha(\bphi)|
=
|\btau^{\prime\prime}\bupsilon\alpha(\bphi)| + |\alpha(\bupsilon)|
=
|\bpsi| + |\bupsilon|
>
|\bpsi|
\geq
|\btau^\prime|
\,
.
\end{equation}
However, since $\bsigma^\prime\in L_\supseteq$,
the prefix and the suffix of $\bsigma^\prime$
contain distinct letters
and they satisfy
$[\btau^{\prime\prime}\bupsilon\alpha(\bupsilon)\alpha(\bphi)]
\subseteq [\btau^\prime]$,
this contradicts \eqref{eq: len comp}.

Since $L_\supseteq$ is the reversal of the language $L_\subseteq$
from~\Cref{e: subset},
the class of languages accepted by
one-register alternating finite-memory automata is not closed
under reversal.\footnote{
Actually, the non-closure under reversal has been already established in~\cite[Exercise 40]{Bojanczyk19}.
However, the proof there is rather involved
and relies on simulating runs of Minsky machines.}
\end{example}

\begin{example}
\label{e: concat}
The concatenation
$L_{\text{co}}=
L_{\text{last}}
\cdot
L_{\text{first}} $
of the languages from Examples~\ref{e: l first} and~\ref{e: l last}
is not accepted by a one-register alternating finite-memory automaton.
Consequently, this class of languages is not closed under concatenation.

For a word $w=\sigma_1\sigma_2\cdots\sigma_m$, a cut after position
$j$, where $1\leq j<m$, witnesses $w\in L_{\mathrm{co}}$ exactly when
\[
\sigma_j\notin[\sigma_1\sigma_2\cdots\sigma_{j-1}]
\quad\hbox{and}\quad
\sigma_{j+1}\notin[\sigma_{j+2}\sigma_{j+3}\cdots\sigma_m].
\]
Thus
the complement language
$\overline{L_{\text{co}}}$ consists
of all words $\sigma_1\sigma_2\cdots\sigma_m$
such that for all $j= 1,2,\ldots,m-1$ 
either $\sigma_{i}\in [\sigma_1\sigma_2\cdots\sigma_{i-1}]$ or
$\sigma_{i+1}\in [\sigma_{i+2}\sigma_{i+3}\cdots\sigma_{m}]$
(cf.~\cite{Okhotin05}).

Consider the auxiliary language
\[
K=\{\#\bpsi\#\#\bomega\#:
\bpsi,\bomega\in L_{\mathrm{diff}},\quad
[\bpsi],[\bomega]\subseteq\Sigma\setminus\{\#\}\}.
\]
A one-register alternating automaton accepts $K$: finite control
checks the four delimiters, and the distinctness test of
\Cref{e: diff} is applied separately to the two blocks.
We claim that
\begin{equation*}
\label{eq: l=}
K\cap\overline{L_{\mathrm{co}}}
=K_=:=\{\#\bpsi\#\#\bomega\#\in K:[\bpsi]=[\bomega]\}.
\end{equation*}
If the blocks have equal contents, cuts before letters of $\bpsi$
fail the second condition, and cuts after letters of $\bomega$
fail the first. All remaining cuts fail because of repeated
delimiters. Conversely, a letter in $[\bpsi]\setminus[\bomega]$
gives a successful cut immediately before it, and a letter in
$[\bomega]\setminus[\bpsi]$ gives one immediately after it.

The language $K_=$ violates \Cref{t: pumping_lemma}.
Indeed, suppose that it is accepted by an automaton $\bA$, and choose
$\bpsi\in L_{\mathrm{diff}}$ avoiding $\#$ with
$|\bpsi|>N_{\sba}$.
In the word $\#\bpsi\#\#\bpsi\#$, every decomposition supplied
by the theorem has its entire first block and the following
pair of delimiters in $\btau$.
Every word of $K_=$ with this fixed prefix has length
$2|\bpsi|+4$, whereas pumping once increases the length by
$|\bupsilon|>0$. This is impossible.
If $L_{\mathrm{co}}$ were recognizable, closure under Boolean
operations would make $K_=$ recognizable, giving the contradiction.
\end{example}

\begin{example}
\label{l: kleene nonclosure}
Choose distinct delimiters $\#$ and $\$$, and let
\[
B=
\{\$\bomega:\bomega\in L_{\mathrm{last}},\ \$\notin[\bomega]\}
\cup
\{\bpsi\$:\bpsi\in L_{\mathrm{first}},\ \$\notin[\bpsi]\}.
\]
This language is recognizable: the automaton checks that $\$$
occurs exactly once, at an endpoint, and applies the corresponding
test to the remaining word.

Let $H$ consist of words $\$u\$$, where
\[
u=\#\bpsi\#\bomega\#,
\qquad
\bpsi,\bomega\in L_{\mathrm{diff}},
\qquad
\mbox{and}
\qquad
[\bpsi],[\bomega]\subseteq\Sigma\setminus\{\#,\$\}.
\]
To recognize $H$, finite control checks the delimiter pattern.
Within each block, the automaton uses the construction
of $L_{\mathrm{diff}}$ from~\Cref{e: diff}, stopping each checking branch at the next $\#$.
Thus it checks distinctness separately in the two blocks.

Since every word of $B$ contains exactly one $\$$, a word
$\$u\$\in H$ belongs to $B^\ast$ exactly when
$u=vw$ with $v\in L_{\mathrm{last}}$ and
$w\in L_{\mathrm{first}}$.
Such a cut exists precisely when $[\bpsi]\neq[\bomega]$.
Indeed, a letter appearing only in $\bpsi$ permits a cut
just before it, and one appearing only in $\bomega$ permits
a cut just after it. If the blocks have equal contents,
cuts before letters of $\bpsi$ or after letters of $\bomega$
fail because those letters recur in the other block;
all remaining cuts fail because the delimiters repeat.

Consequently, $H\cap\overline{B^\ast}$ consists exactly of
the words above with $[\bpsi]=[\bomega]$.
If $B^\ast$ were recognizable, so would this intersection.
However, choosing $\bomega=\bpsi$ with $|\bpsi|>N_{\sba}$
contradicts~\Cref{t: pumping_lemma}: pumping preserves the
first block and increases the total length, whereas
equal contents and distinctness force the second block
to have the same length as the first.
Consequently,
the class of languages accepted by one-register alternating
finite-memory automata is not closed under Kleene star.
\end{example}

\subsection{Semi-linearity}
\label{s: semi linearity}

In this section,
we show that the set of lengths of words
accepted by a one-register alternating finite-memory automaton
is semi-linear
and can be effectively described.
The proof uses the pumping lemma and the decidability of emptiness
from~\cite{GenkinKP14}.
The argument is general and is not limited to languages over infinite alphabets:
it can be applied to any computational model
that possesses similar properties of pumping and periodicity.

\begin{definition}
\label{d: semilinear}
A set of nonnegative integers is {\em linear} if it is of
the form $\{ a + i b : i \in \N \}$,
for some nonnegative integers $a$ and $b$.
A set of nonnegative integers is {\em semi-linear} if it is
a finite union of linear sets and
a language $L$ is called semi-linear if the set
\[
\left| L \right|
=
\left\{ \left| \bsigma \right| : \bsigma \in L \right\}
\]
is semilinear.
\end{definition}

\begin{remark}
\label{r: natural density}
Every linear set $T=\{ a + i b : i \in \N \}$
is either of cardinality one (if $b=0$) or
is of positive natural density
$\frac{\textstyle 1}{\textstyle b}$, otherwise.
Thus, every semi-linear set
is either finite or has positive natural density.
\end{remark}

\begin{theorem}
\label{t: semi linearity}
Let $\bA$ be a one-register alternating finite-memory automaton.
Then, the set $\left| L(\bA ) \right|$
is semi-linear and can be described effectively.
\end{theorem}

\begin{lemma}
\label{l: extension_lemma}
Let $N_{\sba}\geq 1$ be the constant in
\Cref{t: pumping_lemma}.  If $\bsigma\in L(\bA)$ and
$|\bsigma|>N_{\sba}$, then some word $\bsigma'\in L(\bA)$
has length $|\bsigma'|=|\bsigma|+N_{\sba}!$.
\end{lemma}

\begin{proof}
Choose the decomposition $\bsigma=\btau\bupsilon\bphi$ and
permutation $\alpha$ supplied by \Cref{t: pumping_lemma}.
For every positive integer $k$, the pumped word in
\eqref{eq: pumping} has length $|\bsigma|+k|\bupsilon|$.
Since $1\leq |\bupsilon|\leq N_{\sba}$,
the positive integer $|\bupsilon|$ divides $N_{\sba}!$.
Taking $k=N_{\sba}!/|\bupsilon|$ proves the assertion.
\end{proof}

\begin{proof}[Proof of \Cref{t: semi linearity}]
Let $N=N_{\sba}\geq 1$ and $M=N!$.  For each residue
$i\in\{0,1,\ldots,M-1\}$, define
\[
L_i=\{\bsigma\in L(\bA):
        |\bsigma|>N\text{ and }|\bsigma|\equiv i\pmod M\}.
\]
Each $L_i$ is recognized by a one-register alternating automaton:
the additional condition on length can be checked in finite control.
For every nonempty $L_i$, let $m_i$ be its minimum word length.
\Cref{l: extension_lemma}, applied repeatedly, gives
\[
|L_i|=\{m_i+jM:j\in\N\}.
\]
Indeed, the inclusion from right to left follows by pumping, and
the reverse inclusion follows from minimality of $m_i$ and the
common residue modulo $M$.

The remaining lengths form the finite set
\[
T=\{n\in\{0,1,\ldots,N\}:
           L(\bA)\cap\Sigma^n\neq\emptyset\}.
\]
Consequently,
\[
|L(\bA)|=T\ \cup\!
\bigcup_{\substack{0\leq i<M\\L_i\neq\emptyset}}
\{m_i+jM:j\in\N\},
\]
which is semilinear.  Notice that the set of short \emph{lengths}
is finite; the corresponding language need not be finite over
an infinite alphabet.

This description is effective.  Emptiness for one-register
alternating automata is decidable
\cite[Theorem~1]{GenkinKP14}, so we can compute $T$ and determine
which of the finitely many $L_i$ are nonempty.  For each nonempty
$L_i$, test the lengths $n>N$ with $n\equiv i\pmod M$ in increasing
order, using the same emptiness test for $L(\bA)\cap\Sigma^n$.
The first successful test gives $m_i$.
\end{proof}

The following corollary to \Cref{t: semi linearity} is immediate.

\begin{corollary}
{\em (Cf.~\cite[Theorem 1]{DanieliK25}.)}
For a one-register alternating finite-memory automaton $\bA$,
it is decidable whether $\left| L(\bA ) \right|$ is bounded.\footnote{
This is an infinite alphabet counterpart of {\em finiteness} of
languages over a finite alphabet.}
\end{corollary}

\subsection*{Two-register alternating finite-memory automata}

Fix a distinguished letter $\#$ and consider
\[
L_{\#}=\left\{
(\sigma_1\#)(\sigma_1\sigma_2\#)\cdots
(\sigma_1\cdots\sigma_n\#):
n\geq1,\
\sigma_1\cdots\sigma_n\in L_{\mathrm{diff}},\
\#\notin\{\sigma_1,\ldots,\sigma_n\}
\right\}.
\]
The language $L_{\#}$ is the intersection of the languages
$L_1,\ldots,L_7$ defined below. A block is a maximal sequence
of non-$\#$ letters, including the initial block.
\begin{itemize}
\item
$L_1$ consists of words whose first letter exists and is not $\#$.
\item
$L_2$ consists of words whose last letter is $\#$.
\item
$L_3$ consists of words whose second letter is $\#$.
\item
$L_4$ consists of words in which the letter following every
non-final $\#$ equals the first letter.
\item
$L_5$ consists of words with no repeated letter within a block.
\item
$L_6$ consists of words in which, whenever non-$\#$ letters
$a,b$ occur consecutively, every later occurrence of $a$
is immediately followed by $b$.
\item
$L_7$ consists of words in which, whenever a non-$\#$ letter
$a$ is followed by a non-final $\#$, its next occurrence
exists and is immediately followed by $b\#$ for some $b\neq\#$.
\end{itemize}

Every word of $L_{\#}$ satisfies these conditions.
Conversely, $L_1$--$L_4$ ensure that the first block contains
one letter and every block starts with that letter.
Condition $L_6$ forces each subsequent block to contain the
preceding block as a prefix. Conditions $L_5$ and $L_7$
then force it to extend that prefix by exactly one new letter.
Induction therefore gives precisely the words of $L_{\#}$.

Languages $L_1$, $L_2$, and $L_3$ require only finite control.
A one-register deterministic automaton checks $L_4$ by storing
the first letter and comparing it with the letter after each
non-final $\#$.

The complement of $L_5$ is recognized by a one-register
nondeterministic automaton that selects and stores a non-$\#$
letter, then detects its repetition before the next separator.

The complement of $L_6$ is recognized with two registers:
select consecutive non-$\#$ letters $a,b$, store them, and
find a later occurrence of $a$ not immediately followed by $b$.
This includes an occurrence at the end of the word.
If $a=b$, one register suffices.

The complement of $L_7$ is recognized with one register:
select and store $a$, verify that it is followed by a non-final
$\#$, and search for its next occurrence, starting immediately
after that separator. Accept if no such occurrence exists,
or if it is not followed by a non-$\#$ letter and then $\#$.
Accepting states in the waiting modes detect missing letters
at the end of the word.

By Boolean closure, $L_{\#}$ is therefore accepted by a
two-register alternating finite-memory automaton.
In fact, the violations can be recognized nondeterministically,
so a universal two-register automaton suffices.

However,
\[
|L_{\#}|=\left\{\frac{n^2+3n}{2}:n\geq1\right\}.
\]
This infinite set has natural density zero and hence, by
\Cref{r: natural density}, is not semilinear.
Thus \Cref{t: pumping_lemma} does not extend to alternating
finite-memory automata with two or more registers: its pumping
property would imply semilinearity of the length set.

\section{Proof of
\texorpdfstring{\Cref{t: pumping_lemma}}{the alternating pumping lemma}}
\label{s: pf roadmap}

The proof of~\Cref{t: pumping_lemma} proceeds in four steps.
\begin{enumerate}
\item
We analyse an accepting run backwards.
This reversal yields a comparable pair of profiles within a
computably bounded suffix, with the later configuration set
compared to the earlier one.

\item
We introduce $\preceq_\alpha$ to
compare configuration sets relative to a permutation $\alpha$.
We prove that this comparison allows accepting continuations
to be simulated after renaming their letters by $\alpha$.

\item
We turn the profile comparison into a
comparison under $\preceq_\alpha$ by constructing a suitable
permutation $\alpha$.

\item
Finally, we iterate the simulation
to obtain pumping. The key is the direction supplied by
reversal: a continuation from the earlier configuration set
can be replayed from the later one.
\end{enumerate}

\subsection{Trace sequences}
\label{s: sufficient}

Fix a one-register alternating finite-memory automaton $\bA$
with state set $S$, accepting-state set $F$, and distinguished
alphabet $\Delta$.

\begin{definition}
\label{d: configuration-profiles}
For a finite configuration set $C$, a letter $a\in\Sigma$,
and a letter set $\Lambda\subseteq\Sigma$, define
\begin{equation*}
\label{eq: profile-notation}
\begin{aligned}
S^a(C)&=\{s\in S:(s,a)\in C\},&
S^\Lambda(C)&=\bigcup_{a\in\Lambda}S^a(C),\\
\Sigma(C)&=\{a\in\Sigma:S^a(C)\neq\emptyset\}.
\end{aligned}
\end{equation*}
Let $\mathcal Q=2^S\setminus\{\emptyset\}$ and
$r=|\mathcal Q|=2^{|S|}-1$.
For finite $\Lambda$, the \emph{profile vector} is
\begin{equation}
\label{eq: profile-vector}
v_\Lambda(C)\in\N^{\mathcal Q},\qquad
v_\Lambda(C)(Q)
=|\{a\in\Lambda:S^a(C)=Q\}|
\quad(Q\in\mathcal Q).
\end{equation}
Vectors are ordered component-wise:
\begin{equation}
\label{eq: vector-order}
f\leq g
\quad\Longleftrightarrow\quad
f(Q)\leq g(Q)\text{ for every }Q\in\mathcal Q.
\end{equation}
A configuration set $C$ is \emph{accepting} if
$S^\Sigma(C)\subseteq F$.
\end{definition}

\begin{remark}
Since distinguished letters never occur in registers, for every finite set of configurations $C$,
$\Sigma(C)\cap\Delta=\emptyset$.
\end{remark}

\begin{definition}
\label{d: trace}
Let $\bC=(C_h)_{0\leq h\leq m}$ be a run on a word
$\bsigma$ of length $m$,
\begin{equation*}
\label{eq: run}
C_0
\trans{\sigma_1}
C_1
\trans{\sigma_2}
C_2
\trans{\sigma_3}
\cdots
\trans{\sigma_m}
C_m
\,
.
\end{equation*}

For $0\leq i\leq m$, let
$\Sigma_i=\{\sigma_h:m-i<h\leq m\}$,
the set of letters occurring in the last $i$ positions.
Notice
$|\Sigma_i|\leq i$
and $\Sigma_i\subseteq \Sigma_j$ for $i<j$.

For $i=0,1,\ldots,m$ define
\begin{equation}
\label{eq: trace-entries}
T_i=S^{\Sigma\setminus\Sigma_i}(C_{m-i})\qquad
\mbox{and}
\qquad
f_i=v_{\Sigma_i}(C_{m-i})\, .
\end{equation}

The
\emph{trace reversal sequence} of
the run $\bC$ on $\bsigma$ is
\begin{equation*}
\label{eq: trace-sequences}
\operatorname{tr}(\bsigma,\bC)
=\bigl((T_i,f_i)\bigr)_{0\leq i\leq m}.
\end{equation*}
\end{definition}

\begin{definition}
\label{d: admissible-sequence}
A sequence $((T_i,f_i))_{0\leq i<\ell}$ is \emph{admissible} if
\begin{equation}
\label{eq: admissible-sequence}
T_i\subseteq S,\qquad
f_i\in\N^{\mathcal Q},\qquad
f_i(Q)\leq i
\quad(0\leq i<\ell,\ Q\in\mathcal Q).
\end{equation}
Trace entries are partially ordered by
\begin{equation}
\label{eq: trace-order}
(T,f)\leq_{\mathrm{tr}}(U,g)
\quad\Longleftrightarrow\quad
T=U\text{ and }f\leq g,
\end{equation}
using~\eqref{eq: vector-order}.
A \emph{comparable pair} consists of indices $i<j$ with
$(T_i,f_i)\leq_{\mathrm{tr}}(T_j,f_j)$.
An admissible sequence is \emph{bad} if it has no comparable pair, otherwise it is called \emph{good}.
\end{definition}

Every trace reversal sequence
from~\Cref{d: trace} is admissible.

\begin{lemma}
\label{t: na}
There is a computable positive integer $N$, depending only on
$|S|$, such that every admissible sequence
of length at least $N$ is good.
\end{lemma}

\begin{proof}
Consider the prefix tree $\mathcal T$ of bad admissible sequences
from~\Cref{d: admissible-sequence}, rooted at the empty sequence.
By~\eqref{eq: admissible-sequence}, a node at depth $\ell$
has at most $2^{|S|}(\ell+1)^r$ children,
which can be effectively enumerated.

An infinite branch would contain infinitely many entries with
the same first component. Dickson's lemma~\cite{Dickson1913}
would give a comparable pair of their vectors, contradicting
badness under~\eqref{eq: trace-order}.
Thus $\mathcal T$ is finite by K\"onig's infinity
lemma~\cite{Konig26}.

Enumerate its levels until the first empty level $N$.
Then no bad admissible sequence has length $N$ or greater,
and the construction depends only on $|S|$.
\end{proof}

\subsection{Configuration comparison}
\label{s: sufficient 1}

\begin{definition}
\label{d: part_ord}
Let $D_1,D_2$ be finite sets of configurations,
$\Gamma_1,\Gamma_2\subseteq \Sigma$, and
let $\alpha$ be a $\Delta$-permutation of $\Sigma$.
We write
\begin{equation}
\label{eq: preceq}
D_1,\Gamma_1 \preceq_\alpha D_2,\Gamma_2,
\end{equation}
if
\begin{romanenumerate}
\item
\label{eq: comparison-domain}
\label{cl: 1}
$\Gamma_1\subseteq\alpha\left(\Gamma_2\right)$;
\item
\label{eq: comparison-support}
\label{cl: 2}
$\alpha\left(\Gamma_2\right)
\cap
\Sigma\left(D_1\right)\subseteq\Gamma_1$;
\item
\label{eq: comparison-relevant}
\label{cl: 3}
for all $\sigma \in \Gamma_1$,
$S^{\sigma}
\left( D_1 \right)
\subseteq
S^{\alpha^{-1}(\sigma)} \left (D_2 \right)$;
and
\item
\label{eq: comparison-outside}
\label{cl: 4}
$S^{\Sigma\setminus\Gamma_1}\left(D_1\right)\subseteq
S^{\Sigma\setminus\Gamma_2}\left(D_2\right)$.
\end{romanenumerate}
\end{definition}

The intuition behind this definition
arises from the need to embed $D_1$,
associated with $\Gamma_1$, into $D_2$ associated with $\Gamma_2$.
This embedding is mediated by the $\Delta$-permutation $\alpha$,
that plays a central role in the pumping construction
described in the proof overview.
Even though, the actual embedding is performed via $\alpha^{-1}$,
we state the relation in the terms of $\alpha$,
because $\alpha$ is the permutation that will act on the pattern in pumping.

Intuitively, clause (\ref{cl: 1})
ensures that the embedding is well-defined,
while
clause (\ref{cl: 2}) prevents 'overshooting'
the intended domain.
Clause (\ref{cl: 3}) guarantees that the
structural information is preserved under
the symbol transformation, and
clause (\ref{cl: 4}) ensures consistency
of symbols outside the specified domains.

\begin{remark}
\label{r: acc_set}
By clauses (\ref{cl: 3}) and (\ref{cl: 4})
of the above definition,~\eqref{eq: preceq}
implies
\begin{equation*}
\label{eq: acceptance-transfer}
S^\Sigma(D_1)\subseteq S^\Sigma(D_2).
\end{equation*}
In particular,
if $D_2$ is a set of accepting configurations, then so is $D_1$.
\end{remark}

\begin{lemma}
\label{l: suff 1}
Let $D_1,D_2,E_2$
be finite configuration sets, let $\Gamma\subseteq\Sigma$,
and let $\alpha$ be a $\Delta$-permutation.
Suppose
\begin{equation}
\label{eq: simulation-assumptions}
D_1,\alpha(\Gamma)
\preceq_\alpha D_2,\Gamma,
\qquad
D_2\transword{\omega}E_2,
\qquad [\bomega]\subseteq\Gamma.
\end{equation}
There is a finite configuration set $E_1$ such that
\begin{equation}
\label{eq: simulation-result}
D_1\transword{\alpha(\omega)}E_1,
\qquad
E_1,\alpha(\Gamma)\preceq_\alpha E_2,\Gamma.
\end{equation}
\end{lemma}

\begin{proof}
We first establish the one-step case.
Suppose $D_2\trans{\sigma}E'_2$,
$\sigma\in\Gamma$, and
${D_1,\alpha(\Gamma)\preceq_\alpha D_2,\Gamma}$.
Choose successor sets
\begin{equation}
\label{eq: chosen-successors}
E_{2,d}\in\mu^{\sbc}(d,\sigma)\quad(d\in D_2),
\qquad
E'_2=\bigcup_{d\in D_2}E_{2,d}.
\end{equation}
For each $c=(s,a)\in D_1$,
conditions~\eqref{eq: comparison-relevant}
and~\eqref{eq: comparison-outside} give
$d_c=(s,b_c)\in D_2$ with
\begin{equation}
\label{eq: matched-registers}
b_c=\alpha^{-1}(a)\text{ if }a\in\alpha(\Gamma),
\qquad
b_c\notin\Gamma\text{ otherwise}.
\end{equation}
Since $\sigma\in\Gamma$, equation~\eqref{eq: matched-registers} gives
$b_c=\sigma$ if and only if $a=\alpha(\sigma)$.
Therefore
\begin{equation}
\label{eq: local-renaming}
\rho_c:\{b_c,\sigma\}\longrightarrow\{a,\alpha(\sigma)\},
\qquad
\rho_c(b_c)=a,\quad \rho_c(\sigma)=\alpha(\sigma)
\end{equation}
is a bijection between the displayed finite sets.
Since $a,b_c\notin\Delta$ and $\alpha$ fixes $\Delta$,
this bijection fixes every distinguished letter in its
domain and maps non-distinguished letters to
non-distinguished letters.
It therefore extends to a $\Delta$-permutation of $\Sigma$,
which we also denote by $\rho_c$.
By~\Cref{p: permutation word},~\eqref{eq: chosen-successors},
and~\eqref{eq: local-renaming}
we obtain that
\begin{equation}
\label{eq: simulated-successors}
E_{1,c}=\rho_c\left(E_{2,d_c}\right)
\in \mu^{\sbc}(\rho_c(d_c),\rho_c(\sigma))
=
\mu^{\sbc}(c,\alpha(\sigma)),
\qquad
E'_1=\bigcup_{c\in D_1}E_{1,c}.
\end{equation}
By the definition of $\mu^{\sbc}$, every register letter
occurring in $E_{2,d_c}$ belongs to $\{b_c,\sigma\}$.
For $x\in\{b_c,\sigma\}$,
equations~\eqref{eq: matched-registers}--\eqref{eq: local-renaming}
imply
\begin{equation}
\label{eq: successor-comparison}
\begin{aligned}
\rho_c(x)\in\alpha(\Gamma)
&\Longrightarrow x\in\Gamma
\text{ and }\rho_c(x)=\alpha(x),\\
\rho_c(x)\notin\alpha(\Gamma)
&\Longrightarrow x=b_c\notin\Gamma.
\end{aligned}
\end{equation}
Together with~\eqref{eq: chosen-successors}
and~\eqref{eq: simulated-successors},
equation~\eqref{eq: successor-comparison} verifies
\eqref{eq: comparison-relevant} and~\eqref{eq: comparison-outside}
for $E'_1,\alpha(\Gamma)\preceq_\alpha E'_2,\Gamma$.
Conditions~\eqref{eq: comparison-domain}
and~\eqref{eq: comparison-support} are immediate.
We have therefore proved
\begin{equation}
\label{eq: one-step-simulation}
D_1\trans{\alpha(\sigma)}E'_1,
\qquad
E'_1,\alpha(\Gamma)\preceq_\alpha E'_2,\Gamma.
\end{equation}

We prove the lemma by induction on $|\bomega|$.
If $|\bomega|=0$, then $E_2=D_2$ and $E_1=D_1$ satisfies
\eqref{eq: simulation-result} by~\eqref{eq: simulation-assumptions}.
Suppose the assertion holds for words of length $n$, and let
$\bomega=\sigma\bomega'$ have length $n+1$.
Factor the given run as
$D_2\trans{\sigma}E'_2
\transword{\omega'}E_2$.
Apply~\eqref{eq: one-step-simulation} to obtain $E'_1$.
Since $[\bomega']\subseteq\Gamma$, the induction hypothesis,
applied to $E'_1,E'_2$ and the remaining run, gives
$E'_1\transword{\alpha(\omega')}E_1$
and the required final comparison.
Concatenation proves~\eqref{eq: simulation-result}.
\end{proof}

\subsection{From profile counts to a permutation}
\label{s: strong}

\begin{lemma}
\label{l: suff 2}
Let $D_1,D_2$ be finite configuration sets and let
$\Gamma_1\subseteq\Gamma_2$ be finite letter sets satisfying
\begin{equation}
\label{eq: profile-comparison}
v_{\Gamma_1}(D_1)
\leq v_{\Gamma_2}(D_2),\qquad
S^{\Sigma\setminus\Gamma_1}(D_1)
=S^{\Sigma\setminus\Gamma_2}(D_2).
\end{equation}
There are a $\Delta$-permutation $\alpha$ and a letter set
$\Lambda\subseteq\Sigma$ such that
\begin{equation}
\label{eq: permutation-result}
\Gamma_2\subseteq\Lambda,\qquad
\alpha(\Lambda)\subseteq\Lambda,\qquad
D_1,\alpha(\Lambda)\preceq_\alpha D_2,\Lambda.
\end{equation}
\end{lemma}

\begin{proof}
Adjoining $\Delta$ to $\Gamma_1$ and $\Gamma_2$ preserves
\eqref{eq: profile-comparison}, since distinguished letters never
occur in registers. We may therefore assume
$\Delta\subseteq\Gamma_1$.

By~\eqref{eq: profile-vector} and~\eqref{eq: profile-comparison},
for each nonempty state profile the letters carrying that profile
in $\Gamma_1$ inject into the letters carrying the
same profile in $\Gamma_2$.
These injections have disjoint domains and images.
Extend their union by the identity on $\Delta$, then inject
the remaining letters of $\Gamma_1$ into unused letters of
$\Gamma_2$.
This is possible because
$|\Gamma_1\setminus\Delta|
\leq|\Gamma_2\setminus\Delta|$.
Every remaining domain letter has an empty profile, so the
resulting injection $\iota$ satisfies
\begin{equation}
\label{eq: profile-injection}
\iota(\delta)=\delta\quad(\delta\in\Delta),
\qquad
S^a(D_1)\subseteq S^{\iota(a)}(D_2)\quad(a\in\Gamma_1).
\end{equation}

Choose disjoint countably infinite letter sets
$\Theta_+,\Theta_-$ outside
$\Gamma_2\cup\Sigma(D_1)\cup\Sigma(D_2)$.
Choose bijections
\begin{equation*}
\label{eq: fresh-bijections}
\begin{aligned}
\beta_+&:(\Gamma_2\setminus\iota(\Gamma_1))\cup\Theta_+
\longrightarrow\Theta_+,\\
\beta_-&:\Theta_-\longrightarrow
(\Gamma_2\setminus\Gamma_1)\cup\Theta_-.
\end{aligned}
\end{equation*}
These exist because their domains and ranges are countably infinite.
Define
\begin{equation}
\label{eq: constructed-permutation}
\alpha(x)=
\begin{cases}
\iota^{-1}(x),&x\in\iota(\Gamma_1),\\
\beta_+(x),&x\in(\Gamma_2\setminus\iota(\Gamma_1))\cup\Theta_+,\\
\beta_-(x),&x\in\Theta_-,\\
x,&x\notin\Gamma_2\cup\Theta_+\cup\Theta_-.
\end{cases}
\end{equation}
The domain sets in~\eqref{eq: constructed-permutation} partition
$\Sigma$, as do their images; see also~\Cref{fig: permutation-construction}.
Hence $\alpha$ is a permutation, and
\eqref{eq: profile-injection} shows that it fixes $\Delta$.
For $\Lambda=\Gamma_2\cup\Theta_+$,
equation~\eqref{eq: constructed-permutation} gives
\begin{equation}
\label{eq: invariant-letter-set}
\alpha(\Lambda)=
\alpha(\Gamma_2)\cup \alpha(\Theta_+)
=
\Gamma_1\cup\Theta_+\subseteq\Lambda.
\end{equation}

It remains to show that $D_1,\alpha(\Lambda)\preceq_\alpha D_2,\Lambda$.
Conditions~\eqref{eq: comparison-domain}
and~\eqref{eq: comparison-support}
are $\alpha(\Lambda)\subseteq \alpha(\Lambda)$
and $\alpha(\Lambda)\cap\Sigma(D_1)\subseteq \alpha(\Lambda)$
which are immediate.
For condition~\eqref{eq: comparison-relevant}
we need to show that for all $\sigma\in \alpha(\Lambda)$,
$S^{\sigma}(D_1)\subseteq S^{\alpha^{-1}(\sigma)}(D_2)$.
If $\sigma\in \Gamma_1$, then $\alpha^{-1}(\sigma)=\iota(\sigma)$ and the inclusion follows from~\eqref{eq: profile-injection}.
If $\sigma\in \Theta_+$, $S^{\sigma}(D_1)$
is empty and the inclusion holds trivially.
Finally, condition~\eqref{eq: comparison-outside}
follows from~\eqref{eq: profile-comparison}
because letters from $\Theta_+$ and $\Theta_-$
are not in $\Sigma(D_1)\cup\Sigma(D_2)$
therefore,
\begin{equation}
\label{eq: outside-equality}
S^{\Sigma\setminus\alpha(\Lambda)}(D_1)
=S^{\Sigma\setminus\Gamma_1}(D_1)
=S^{\Sigma\setminus\Gamma_2}(D_2)
=S^{\Sigma\setminus\Lambda}(D_2).
\end{equation}
Together with~\eqref{eq: invariant-letter-set}
and $\Gamma_2\subseteq\Lambda$, this proves~\eqref{eq: permutation-result}.
\end{proof}

\begin{figure}[htbp]
\centering
\begin{tikzpicture}[
  font=\small,
  map/.style={-{Stealth[length=2mm]},semithick},
  fresh/.style={draw,circle,minimum size=12mm,inner sep=0pt}
]
  \draw[semithick]
    (-2.4,0) ellipse[x radius=1.55cm,y radius=1.85cm];
  \draw[semithick]
    ( 2.4,0) ellipse[x radius=1.55cm,y radius=1.85cm];

  \draw[densely dashed] (-3.95,0) -- (-0.85,0);
  \draw[densely dashed] ( 0.85,0) -- ( 3.95,0);

  \node at (-2.4, 2.15) {$\Gamma_2$};
  \node at ( 2.4, 2.15) {$\Gamma_2$};
  \node at (-2.4,-2.15) {codomain};
  \node at ( 2.4,-2.15) {domain};

  \node at (-2.4, 0.85) {$\Gamma_1$};
  \node at (-2.4,-0.85) {$\Gamma_2\setminus\Gamma_1$};
  \node at ( 2.4, 0.85) {$\iota(\Gamma_1)$};
  \node at ( 2.4,-0.85)
    {$\Gamma_2\setminus\iota(\Gamma_1)$};

  \draw[map]
    (1.03,0.85) -- node[above] {$\iota^{-1}$} (-1.03,0.85);

  \node[fresh] (minus) at (-5.7,-0.85) {$\Theta_-$};
  \node[fresh] (plus)  at ( 5.7,-0.85) {$\Theta_+$};

  \draw[map]
    (minus.east) -- node[below] {$\beta_-$} (-3.77,-0.85);
  \draw[map]
    (3.77,-0.85) -- node[below] {$\beta_+$} (plus.west);

  \path[map]
    (minus) edge[loop above,min distance=12mm]
    node[above] {$\beta_-$} (minus);
  \path[map]
    (plus) edge[loop above,min distance=12mm]
    node[above] {$\beta_+$} (plus);
\end{tikzpicture}

\caption{The construction of the profile-preserving permutation $\alpha$.}
\label{fig: permutation-construction}
\end{figure}

\subsection{Completing the proof}
\label{s: completing the proof}

\begin{proof}[Proof of \Cref{t: pumping_lemma}]
Let $N_{\sba}=N$ from~\Cref{t: na}, and fix an accepting run
$(E_h)_{0\leq h\leq m}$ on $\bsigma$, where $m>N$.
Its trace reversal sequence from~\Cref{d: trace} is admissible.
Applying~\Cref{t: na} to its first $N$ entries gives
$0\leq i<j<N$ with
$(T_i,f_i)\leq_{\mathrm{tr}}(T_j,f_j)$.

Choose
${\btau=\sigma_1\sigma_2\cdots\sigma_{m-j}}$,
${\bupsilon=\sigma_{m-j+1}\sigma_{m-j+2}\cdots\sigma_{m-i}}$,
and
${\bphi=\sigma_{m-i+1}\sigma_{m-i+2}\cdots\sigma_m}$.
Empty and singleton index ranges denote the empty word
and the corresponding letter, respectively.
Notice
$|\bupsilon|=j-i>0$ and
$|\bupsilon\bphi|=j<N_{\sba}. $
Define
\begin{equation}
\label{eq: pumping-decomposition}
D_1=E_{m-i},\qquad D_2=E_{m-j},\quad
\Gamma_1=[\bphi],\qquad \Gamma_2=[\bupsilon\bphi].
\end{equation}
Hence
\begin{equation}
\label{eq: selected-run}
E_0\transword{\tau}D_2
\transword{\upsilon}D_1
\transword{\varphi}E_m.
\end{equation}

By~\eqref{eq: trace-entries} and~\eqref{eq: trace-order},
the sets in~\eqref{eq: pumping-decomposition} satisfy
\eqref{eq: profile-comparison}.
Since $\Gamma_1\subseteq\Gamma_2$, \Cref{l: suff 2} gives
a $\Delta$-permutation $\alpha$ and a letter set $\Lambda$ with
\begin{equation}
\label{eq: pumping-comparison}
D_1,\alpha(\Lambda)\preceq_\alpha D_2,\Lambda,
\qquad
[\bupsilon\bphi]\subseteq\Lambda,\qquad
\alpha(\Lambda)\subseteq\Lambda.
\end{equation}

Define continuations recursively by
\begin{equation}
\label{eq: pumping-continuations}
\bomega_0=\bupsilon\bphi,\qquad
\bomega_{k+1}=\bupsilon\alpha(\bomega_k)
\quad(k\geq0).
\end{equation}
We prove by induction on $k$ that
$[\bomega_k]\subseteq\Lambda$ and that $D_2$ has an accepting
run on $\bomega_k$.
The case $k=0$ follows from~\eqref{eq: selected-run}
and~\eqref{eq: pumping-comparison}.

Suppose the assertion holds at $k$.
By~\Cref{l: suff 1,r: acc_set} and
\eqref{eq: pumping-comparison}, $D_1$ has an accepting run
on $\alpha(\bomega_k)$.
Prepending the transition
$D_2\transword{\upsilon}D_1$
from~\eqref{eq: selected-run} gives an accepting run from
$D_2$ on $\bomega_{k+1}$.
Moreover,
$[\bomega_{k+1}]
=[\bupsilon]\cup\alpha([\bomega_k])\subseteq\Lambda$
by~\eqref{eq: pumping-comparison}.
This completes the induction.

Thus the comparison obtained by reversal permits repeated
simulation of the continuation. Prefixing these accepting runs
by the run on $\btau$ in~\eqref{eq: selected-run}, and expanding
\eqref{eq: pumping-continuations}, gives
\begin{equation*}
\label{eq: pumped-word}
\btau\bupsilon\alpha(\bupsilon)\cdots
\alpha^k(\bupsilon)\alpha^k(\bphi)\in L(\bA)
\qquad(k\geq1).
\end{equation*}
\end{proof}

\section{Concluding remarks}
\label{s: remark}

In this paper,
we introduced and proved a pumping-like lemma
for languages over infinite alphabets,
specifically those recognized by one-register alternating
finite-memory automata.
As a key corollary, we established that the set of word lengths
in such languages is semilinear.

The central contribution lies not only
in the results themselves but also
in the underlying technique,
a novel shift in perspective for analyzing transition
systems that are not well-structured
in the forward direction but can be controlled
in reverse.
At reverse index $i$, only letters
in the last $i$ positions need individual profiles. 
This reverse-analysis allows us to recover
a form of well-quasi-ordering by restricting
attention to suffix-dependent symbol
domains,
enabling a form of pumping that
respects the structural symmetries of infinite alphabets.

{\bf Complexity of computing $\bN_{\hspace{-0.15 em}} \sba$}.
Most algorithms for computational models
over unbounded domains
(such as finite-memory automata, Petri nets,
vector addition systems with states)
have extremely high complexity,
often beyond primitive-recursive bounds.
These complexities are typically classified
within the fast-growing hierarchy
$\mathfrak{F}$~\cite{Lob71}.

In our case,
for a fixed number of states $n=|S|$,
bad trace reversal sequences
are controlled bad sequences over $2^n$ copies of $\N^{2^n-1}$.
By~\cite[Proposition~5.2]{FigueiraFSS11},
their length
bound lies in $\mathfrak F_{2^n-1}$.
Therefore the elementary overhead of
tree enumeration yields a computation bound in
$\mathfrak F_{\max\{2,2^n-1\}}$.

Furthermore,
our bad-sequence constant also bounds shortest
accepted words of one-register
alternating finite-memory
automata~\cite[Lemmas~14 and~16]{GenkinKP14}.
Since their nonemptiness is known
to be non-primitive recursive
\cite[Theorem~5.2]{DemriL09},
the function $\bA \mapsto N_\sba$
is not primitive recursive.

{\bf Ordered alphabets}.
A natural question is whether our results
can be extended to any orbit-finite
alphabet \cite{BojanczykKL14}.
For example, alphabets with some order relation.
In that setting, one might consider
replacing the notion of a $\Delta$-permutation
with that of an order-preserving permutation.

Here, the main technical challenge lies
in the construction of
an order-preserving permutation
that generates infinitely many ordered copies of names.
This is generally not possible
for arbitrary orders.

Consider, for example,
the alphabet $\Sigma=(\BN,\leq)$ and the language
\[
L_{ord}=
\{\sigma_1\sigma_2\cdots\sigma_n:
\ \mbox{for all} \ i<j,\sigma_i > \sigma_j\}
\]
that is accepted by a one-register alternating automaton
with order comparisons.
This language is unbounded, yet it contains no
pumpable patterns: for any positive integer $N$,
the word $N \cdot (N-1) \cdots  1$
cannot be pumped while preserving the strict decreasing order.

Nevertheless, if the underlying order is dense and total such as
$(\bQ,\leq)$, Higman's lemma
\cite{Higman52} suggests an approach using ordered sequences of state
profiles.

{\bf Timed automata}.
Our results for register automata also raise a question
about one-clock alternating timed automata
\cite{AlurD94,LasotaW08}.
The connection comes from the translation of timed automata
into register automata over an ordered alphabet
studied in \cite{FigueiraHL16}.

This translation does not allow us to apply our results directly.
The resulting register automata use order comparisons over
$(\bQ,\leq)$, whereas our pumping lemma concerns equality tests.
An extension to ordered alphabets, as suggested above,
would therefore be needed first.
Moreover, the translation is not linear in word length,
so a result about the lengths of the translated words
would not immediately give the same result for timed words.

We leave open whether the sets of word lengths
of languages accepted by one-clock alternating timed automata
are semilinear.

\bibliographystyle{abbrv}
\bibliography{references}

@article{AlurD94,
author       = {Rajeev Alur and
David L. Dill},
title        = {A Theory of Timed Automata},
journal      = {Theor. Comput. Sci.},
volume       = {126},
number       = {2},
pages        = {183--235},
year         = {1994},
url          = {https://doi.org/10.1016/0304-3975(94)90010-8},
doi          = {10.1016/0304-3975(94)90010-8},
bibsource    = {dblp computer science bibliography, https://dblp.org}
}

@article{BojanczykKL14,
author    = {Mikolaj Bojanczyk and Bartek Klin and
S\lawomir Lasota},
title     = {Automata theory in nominal sets},
journal   = {Logical Methods in Computer Science},
volume    = {10},
number    = {3},
pages     = {1--44},
year      = {2014},
}

@article{Bojanczyk19,
title={Slightly infinite sets},
author={Bojanczyk, Miko{\l}aj},
journal={A draft of a book available at {\url {https://www.mimuw.edu.pl/~bojan/paper/atom-book}}},
year={2019}
}

@article{ChandraKS81,
author    = {A.K. Chandra and D.C. Kozen and L.J. Stockmeyer},
title     = {Alternation},
journal   = {Journal of the ACM},
volume    = {28},
number    = {1},
year      = {1981},
pages     = {114--133}
}

@inproceedings{ChenSW16,
author       = {Taolue Chen and
Fu Song and
Zhilin Wu},
editor       = {Jonathan P. Bowen and
Zhiming Liu and
Zili Zhang},
title        = {Formal Reasoning on Infinite Data Values: An Ongoing Quest},
booktitle    = {Engineering Trustworthy Software Systems - Second International School,
{SETSS} 2016, Chongqing, China, March 28 - April 2, 2016, Tutorial
Lectures},
series       = {Lecture Notes in Computer Science},
volume       = {10215},
pages        = {195--257},
year         = {2016},
url          = {https://doi.org/10.1007/978-3-319-56841-6\_6},
doi          = {10.1007/978-3-319-56841-6\_6},
bibsource    = {dblp computer science bibliography, https://dblp.org}
}

@article{ClementeLP22,
author    = {Lorenzo Clemente and Slawomir Lasota and
Radoslaw Pi{\'{o}}rkowski},
title     = {Determinisability of register and timed automata},
journal   = {Logical Methods in Computer Science},
volume    = {18},
PAGES     = {Paper No. 9, 37},
year      = {2022},
}

@incollection{DanieliK25,
author="Danieli, Yoav
and Kaminski, Michael",
editor="Meer, Klaus
and Rabinovich, Alexander
and Ravve, Elena
and Villaveces, Andr{\'e}s",
title="Bounded Languages Over Infinite Alphabets",
bookTitle="Model Theory, Computer Science, and Graph Polynomials: Festschrift in Honor of Johann A. Makowsky",
year="2025",
publisher="Springer Nature Switzerland",
address="Cham",
pages="193--213",
isbn="978-3-031-86319-6",
doi="10.1007/978-3-031-86319-6_15",
url="https://doi.org/10.1007/978-3-031-86319-6_15"
}

@Inproceedings{DanieliAfl26,
author    = {Danieli, Yoav},
year      = {2026},
title     = {Pumping Constants for Infinite Alphabets},
editor    = {Hospod\'ar, Michal and Jir\'askov\'a, Galina and Olej\'ar, Viktor and \v{S}ebej, Juraj and Vaszil, Gy\"orgy},
booktitle = {{\rm Proceedings 17th International Conference on}
Automata and Formal Languages,
{\rm Ko\v{s}ice, Slovakia, 7-10th September 2026}},
series    = {Electronic Proceedings in Theoretical Computer Science},
volume    = {451},
publisher = {Open Publishing Association},
pages     = {104-122},
doi       = {10.4204/EPTCS.451.8},
}

@inproceedings{DanieliLICS26,
  author       = {Yoav Danieli},
  editor       = {Claudia Faggian and
                  Joost{-}Pieter Katoen},
  title        = {Star Complexity of Parikh Images of Languages over Infinite Alphabets},
  booktitle    = {41st Annual Symposium on Logic in Computer Science, {LICS} 2026, Lisbon,
                  Portugal, July 20-23, 2026},
  series       = {LIPIcs},
  volume       = {380},
  pages        = {35:1--35:26},
  publisher    = {Schloss Dagstuhl - Leibniz-Zentrum f{\"{u}}r Informatik},
  year         = {2026},
  url          = {https://doi.org/10.4230/LIPIcs.LICS.2026.35},
  doi          = {10.4230/LIPICS.LICS.2026.35},
  bibsource    = {dblp computer science bibliography, https://dblp.org}
}

@article{DemriL09,
author    = {S. Demri and R. Lazi\'{c}},
title     = {{LTL} with the freeze quantifier and register automata},
journal   = {ACM Transactions on Computational Logic},
volume    = {10},
year      = {2009},
note      = {Article 16}
}

@article{Dickson1913,
author    = {L.E. Dickson},
title     = {Finiteness of the odd perfect and primitive abundant numbers
with $n$ distinct prime factors},
journal   = {American Journal of Mathematics },
volume    = {35},
year      = {1913},
pages     = {413--422}
}

@inproceedings{FigueiraFSS11,
title       = {Ackermannian and Primitive-Recursive Bounds with {D}ickson's Lemma},
booktitle   = {2011 IEEE 26th Annual Symposium on Logic in Computer Science},
publisher   = {IEEE},
author      = {Figueira, Diego and Figueira, Santiago and Schmitz,
Sylvain and Schnoebelen, Philippe},
year        = {2011},
pages       = {269--278}
}

@article{FigueiraHL16,
author       = {Diego Figueira and Piotr Hofman and S\lawomir Lasota},
title        = {Relating timed and register automata},
journal      = {Mathematical Structures in Computer Science},
volume       = {26},
pages        = {993--1021},
year         = {2016},
}

@article{GenkinKP14,
author    = {D. Genkin and M. Kaminski and L. Peterfreund},
title     = {A note on the emptiness problem for
alternating finite-memory automata},
journal   = {Theoretical Computer Science},
volume    = {526},
pages     = {97--107},
year      = {2014}
}

@article{Higman52,
title={Ordering by divisibility in abstract algebras},
author={Higman, Graham},
journal={Proceedings of the London Mathematical Society},
volume={3},
number={1},
pages={326--336},
year={1952},
publisher={Oxford University Press}
}

@inproceedings{KaminskiF90,
author    = {M. Kaminski and N. Francez},
title     = {Finite-memory automata},
booktitle = {Proceedings of the 31th Annual IEEE Symposium on
Foundations of Computer Science},
publisher = {IEEE Computer Society Press},
address   = {Los Alamitos, CA},
pages     = {683--688},
year      = {1990}
}

@article{KaminskiF94,
author    = {M. Kaminski and N. Francez},
title     = {Finite-memory automata},
journal   = {Theoretical Computer Science},
volume    = {134},
year      = {1994},
pages     = {329--363},
}

@phdthesis{Kara16,
author       = {Ahmet Kara},
title        = {Logics on data words: Expressivity, satisfiability, model checking},
school       = {Technical University of Dortmund, Germany},
year         = {2016},
url          = {https://hdl.handle.net/2003/35216},
urn          = {urn:nbn:de:101:1-201609231316},
bibsource    = {dblp computer science bibliography, https://dblp.org}
}

@misc{Klin14,
author = {B. Klin},
title = {A pumping lemma for automata with atoms},
year   = {26.03.2014},
howpublished = {\url{https://atoms.mimuw.edu.pl/?p=43/}},
note = "[accessed 08.08.2025]"
}

@inproceedings{KlinlT21,
author    = {Bartek Klin and S\lawomir Lasota and Szymon Toru\'nczyk},
editor    = {Stefan Kiefer and Christine Tasson},
title     = {Nondeterministic and co-Nondeterministic Implies
Deterministic, for Data Languages},
booktitle  = {Foundations of Software Science and
Computation Structures~- 24th
International Conference, {FOSSACS} 2021},
series    = {Lecture Notes in Computer Science},
volume    = {12650},
pages     = {365--384},
publisher = {Springer},
year      = {2021},
}

@article{Konig26,
author    = {D. K\"{o}nig},
title     = {Sur les correspondences multivoques des ensembles},
journal   = {Fundamenta Mathematicae},
volume    = {8},
year      = {1926},
pages     = {114--134}
}

@article{Landau03,
author    = {Edmund Landau},
title     = {\"{U}ber die {M}aximalordung der {P}ermutation
gegbenen {G}rades},
journal   = {Archiv der Mathematik und Physik},
volume    = {5},
year      = {1903},
pages     = {92--103},
}

@article{LasotaW08,
author       = {Slawomir Lasota and
Igor Walukiewicz},
title        = {Alternating timed automata},
journal      = {{ACM} Trans. Comput. Log.},
volume       = {9},
number       = {2},
pages        = {10:1--10:27},
year         = {2008},
url          = {https://doi.org/10.1145/1342991.1342994},
doi          = {10.1145/1342991.1342994},
bibsource    = {dblp computer science bibliography, https://dblp.org}
}

@article{Lob71,
author       = {M. L\"{o}b and S. Wainer},
journal      = {Archive for Mathematical Logic},
number       = {3-4},
pages        = {198--199},
publisher    = {Springer Science and Business Media Llc},
title        = {Hierarchies of Number-Theoretic Functions {I}, {II}:
a correction},
volume       = {14},
year         = {1971}
}

@article{NakanishiTS23,
title     = {Pumping Lemmas for Languages Expressed by
Computational Models with Registers},
author    = {Rindo Nakanishi and Yoshiaki Takata and Hiroyuki Seki},
journal   = {IEICE Transactions on Information and Systems},
year      = {2023},
volume    = {106},
pages     = {284--293},
}

@article{NevenSV04,
author    = {F. Neven and T. Schwentick and V. Vianu},
title     = {Finite state machines for strings over infinite alphabets},
journal   = {ACM Transactions on Computational Logic},
volume    = {5},
year      = {2004},
pages     = {403--435},
}

@article{Okhotin05,
author       = {Alexander Okhotin},
title        = {The dual of concatenation},
journal      = {Theoretical Computer Science},
volume       = {345},
number       = {2-3},
pages        = {425--447},
year         = {2005},
}

@article{RabinS59,
author    = {M. Rabin and D. Scott},
title     = {Finite automata and their decision problems},
journal   = {IBM Journal of Research Development},
volume    = {3},
year      = {1959},
pages     = {114--125},
}

@article{SakamotoI00,
author    = {H. Sakamoto and D. Ikeda},
title     = {Intractability of decision problems for
finite-memory automata},
journal   = {Theoretical Computer Science},
volume    = {231},
year      = {2000},
pages     = {297--308}
}

@inproceedings{Segoufin06,
author    = {Luc Segoufin},
editor    = {Zolt{\'{a}}n {\'{E}}sik},
title     = {Automata and Logics for Words and Trees over an
Infinite Alphabet},
booktitle = {Computer Science Logic, 20th International Workshop,
{CSL} 2006, 15th Annual Conference of the EACSL},
series    = {Lecture Notes in Computer Science},
volume    = {4207},
pages     = {41--57},
publisher = {Springer},
year      = {2006},
}

@article{Tan10,
author    = {T. Tan},
title     = {On pebble automata for data languages with
decidable emptiness problem},
journal   = {Journal of Computer and System Sciences},
volume    = {76},
year      = {2010},
pages     = {778--791},
}

\end{document}